\documentclass[12pt,preprint]{aastex}

\begin{document}

\title{Evidence for a Non-Universal Stellar Initial Mass Function from the Integrated Properties of SDSS Galaxies}

\author{Erik A. Hoversten}
\affil{Department of Physics \& Astronomy, Johns Hopkins University}
\affil{3400 N. Charles Street, Baltimore, MD 21218}
\and
\author{Karl Glazebrook}
\affil{Centre for Astrophysics and Supercomputing, Swinburne University of Technology}
\affil{P.O. Box 218, Hawthorn, VIC 3122, Australia}

\shorttitle{Non-Universal IMF in  SDSS Galaxies}

\begin{abstract}

This paper revisits the classical Kennicutt method for inferring the stellar initial
mass function (IMF) from the integrated light properties of galaxies.  The
large size, uniform high quality data set from the Sloan Digital Sky Survey DR4 is
combined with more in depth modeling and quantitative statistical analysis to
search for systematic IMF variations as a function of galaxy luminosity.
Galaxy H$\alpha$ equivalent widths are compared to a broadband color index to
constrain the IMF.  This parameter space is useful for breaking degeneracies
which are traditionally problematic.  Age and dust corrections are largely
orthogonal to IMF variations.  In addition the effects of metallicity and
smooth star formation history e-folding times are small compared to IMF
variations.  We find that for the sample as a whole the best fitting IMF slope
above 0.5 ${\rm M}_\odot$ is $\Gamma =1.4535$ with a negligible random error
of $\pm 0.0004$ and a systematic error of $\pm 0.1$.  Galaxies brighter than
around ${\rm M}_{r,0.1}=-20$ (including galaxies like the Milky Way which has ${\rm M}_{r,0.1}\sim -21$) are well fit by a universal $\Gamma \sim 1.4$ IMF,
similar to Salpeter, and smooth, exponential star formation histories (SFH).
Fainter galaxies prefer steeper IMFs and the quality of the fits reveal that for these galaxies a universal IMF with smooth SFHs is actually a poor assumption.  Several sources of sample bias are ruled out as the cause of these luminosity dependent IMF variations.  Analysis of bursting SFH models
shows that an implausible coordination of burst times is required to fit a
universal IMF to the ${\rm  M}_{r,0.1}=-17$ galaxies.  This leads to the
conclusions that the IMF in low luminosity galaxies has fewer massive stars,
either by steeper slope or lower upper mass cutoff, and is not universal.

\end{abstract}


\section{Introduction}

A precise measurement of the stellar initial mass function (IMF) and its
functional dependence on environmental conditions would impact astronomy over
a wide range of physical scales.  It would be of great help to theorists in
untangling the mysteries of star formation and it is a key input in spectral
synthesis models used to interpret the observed properties of galaxies both
nearby and in the early universe.

The current question is whether the IMF is \textit{universal}-- the same
regardless of time and environmental conditions.  \citet{Ken98} concisely
states the current understanding of IMF universality.  It is difficult to
believe that the IMF is universal given the diversity of galaxy types,
environments, star formation rates, and populations within galaxies over the
range of observable lookback times.  On the other hand, while IMF measurements
do vary they are all consistent with a universal IMF within measurement errors
and sampling statistics.  The only way to proceed then is to strive for
smaller measurement errors and improving sample sizes.

A definitive theoretical derivation of the IMF does not yet exist.
Theoretical approaches to the IMF usually center around the Jeans mass, $M_J$,
the mass at which a homogeneous gas cloud becomes unstable.  At first the
collapse of a cloud is isothermal and the Jeans mass decreases which leads to
fragmentation of the cloud \citep{Hoyle53}.  Both \citet{Rees76} and
\citet{Low76} suggested that at some point during the cloud collapse the line
cooling opacity becomes high enough that the collapse is no longer isothermal.
At this point the Jeans mass increases and fragmentation stops.  The minimum
Jeans mass is the smallest fragment size at this point and it provides a lower
limit to size of the stars formed.

Authors have calculated Jeans masses and minimum Jeans masses using a variety
of methods.  In the classical derivation of the Jeans mass $M_J \propto
T^{3/2} \rho^{-1/2}$.  \citet{Low76} finds that $M_{J,min} \propto m^{-16/7}
\kappa^{-1/7}$ where $m$ is the mass of gas atoms or molecules and $\kappa$ is
the opacity at the final fragmentation.  More recently turbulence in clouds
has been studied.  \citet{Padoan97} found $M_{J,min} \propto n^{-1/2} T^2
\sigma_v^{-1}$ where $n$ is the number density and $\sigma_v$ is the velocity
dispersion of the gas.  Other investigators have looked at the hierarchical
fractal geometry of molecular clouds, thought to arise from turbulence, as a
generator of the IMF (e.g. \citet{Elmegreen97}).  On a related note
\citet{Adams96} point out that molecular clouds exhibit structure on all
resolvable spacial scales suggesting that as no characteristic density exists for
the clouds neither does a single Jeans mass.  They develop a semi-empirical
model for determining the final masses of stars from the initial conditions of
molecular clouds without invoking Jeans mass arguments and use it to construct
IMF models.  The key components of their model are sound speed and rotation
rate of cloud cores and the idea that stars help determine their final masses
through winds and outflows.
 
But from the beginning the study of the IMF has been driven by measurements.
In 1955 Salpeter was the first to make a measurement of the IMF inferring it
from his observed stellar luminosity function \citep{S55}.  We parameterize
the IMF by:
\begin{equation}
\frac{{\rm d}n}{{\rm d}\log m} \propto \left \{ \begin{array}{ccc} -0.5 & \mbox{for}& 0.1 <m/{\rm M}_\odot < 0.5\\
-\Gamma & \mbox{for}& 0.5 < m/{\rm M}_\odot < 120 \end{array} \right.
\label{eq:imfbaldry}
\end{equation}
following \citet{BG03}.  Salpeter found that $\Gamma = 1.35$.  It is often
overlooked that his measurement only covered masses for which $0.4 \lesssim
m/\rm{M}_\sun \lesssim 10$.  Nonetheless his original measurement is
surprisingly consistent with modern values over a wide range of masses.  The
IMF in equation \ref{eq:imfbaldry} is similar to the Salpeter IMF for $\Gamma =
1.35$.  The difference is that there are fewer stars with masses less than 0.5
${\rm M}_\odot$.  We adopt a two part power law as there is agreement amongst
several authors that there is a change in the IMF slope near 0.5 ${\rm
  M}_\odot$ \citep{Kroupa01}.  The technique we will use is not sensitive to
the IMF at low masses so we assume a constant value in that regime.

Salpeter's idea continues to be used today in IMF measurements of resolved
stellar populations.  The technique can be applied to field stars as well as
clusters.  However Salpeter's method has several inherent limitations.  The
nature of stars presents a challenge.  On the high mass end stars are very
luminous, but live only a few million years, while on the low end stars are
faint but have lifetimes many times longer than the current age of the
universe.  There are very few star clusters which are both young and close
enough to allow us access to the IMF over the full mass range.  In addition,
the main sequence mass-luminosity relationship is a function of age,
metallicity, and speed of rotation in addition to mass.  It is not yet
well-known at the low and high mass extremes.  Unresolved binaries can also
affect the measured IMF \citep{Kroupa01}.  The light from unresolved binaries
is dominated by the more massive of the pair.  As a result the less massive
star is typically not detected which leads to a systematic under-counting of
low mass stars.  $42\% \pm 9\%$ of main sequence M stars \citep{Fischer92} and 43\% of main sequence G stars \citep{Duquennoy91} are primary stars in multiple star systems.  These are both lower limits as some companions may have eluded detection.  As roughly half of stars are in
multiple systems it has a potentially large effect on the observed luminosity
function. 

Except for at the high mass end field stars in the solar neighborhood offer the best statistics for IMF
measurements.  However the solar neighborhood IMF is found to be deficient in
massive stars when compared to other galaxies.  For example the \citet{MS79}
and \citet{Scalo86} solar neighborhood IMFs are rejected by integrated light
approaches, i.e.  \citet{K83}, hereafter K83, \citet{K94} (KTC94), and
\citet{BG03}.

Analysis of individual star clusters can be used to detect IMF variations.  As
methods and data quality can vary between authors comparisons between
individual clusters are difficult.  However \citet{Phelps93} studied eight
young open clusters with the same technique.  On the extremes they measured
$\Gamma =1.06 \pm 0.05$ for NGC 663 and $\Gamma = 1.78 \pm 0.05$ for NGC 581
over masses from around 1 to 12 M$_\sun$.

\citet{Kroupa01} considered the case in which one could have perfect knowledge
of the masses of all stars in a cluster.  Clusters have a finite size so even
with no measurement errors uncertainty arises from sampling the underlying
IMF.  He shows that the observed scatter in IMF power law values above 1
M$_\sun$ can be accounted for by sampling bias for clusters with between
$10^2$ and $10^3$ members (the \citet{Phelps93} clusters have memberships in
this range).  Furthermore, dynamical evolution of clusters can affect
measurements of the IMF.  This can happen by preferentially expelling lower
mass stars from the cluster and by breaking up binary systems within the 
cluster.  In total \citet{Kroupa01} finds that for stars with $m \gtrsim 1{\rm
  M}_\sun$ the spread in the observed IMF power-law slopes in clusters is
around 1 when both binary stars and sampling bias are considered even when the
underlying IMFs are identical.

Stochastic processes can also influence the ability to detect systematic IMF
variations.  O and B stars produce ionizing photons and cosmic rays which
affect the surrounding nebula.  However the probability of creating one of
these massive stars is comparatively small.  If one of these massive stars
happens by chance to form first it may drive up the nebular temperature and
depress the formation of less massive stars compared to regions without
massive stars \citep{Robberto04}.

An alternative to IMF measurements based on the stellar luminosity functions of
resolved stellar populations is to infer the characteristics of stellar
populations from the integrated light of galaxies using spectral synthesis
models.  The advantage of using integrated light techniques is that many of
the problems plaguing IMF investigations of resolved stellar populations are
avoided.  Stochastic effects are washed out over a whole galaxy.  Unresolved
multiple star systems are irrelevant.  The number of observable galaxies is
large and their environments span a much larger range than those of Milky Way
clusters.  Integrated light techniques can be applied to the high redshift
universe.  This creates a strong motivation to develop IMF techniques and test
them for galaxies in the local universe which can later be used to probe
earlier stages of galaxy formation.  The assumption of a universal IMF has a
huge influence on the interpretation of the high redshift universe.  The
reionization of the universe and the Madau plot-- the global star formation
rate of the universe over cosmic history-- are two areas where IMF variations
could impact the current picture of galaxy formation and evolution. 

On the downside conclusions are dependent on the stellar evolution models
used, which are not well constrained at high masses or with horizontal branch
stars at low metallicities.  The biggest problem is that changes in the IMF,
star formation history (SFH) or age of a galaxy model can have similar effects
in the resulting spectral energy distribution (SED).  Any integrated light
technique needs to address these degeneracies.  It is also difficult to probe
the IMF at sub-solar masses using integrated light.

On the level of galaxies the concept of a universal IMF has recently become
more complicated.  A number of recent studies have investigated the effect of
summing the IMF in individual clusters over a galaxy in the presence of power
law star cluster mass functions.  \citet{WK05} argue that the integrated
galaxial IMF will appear to vary as function of galactic stellar mass even if
the stellar IMF is universal.  However \citet{Elmegreen06} argues that the
galaxy wide IMF should not differ from the IMF in individual clusters based on
analytical arguments and Monte Carlo simulations.  Our approach can only
measure the IMF averaged over whole galaxies and cannot address this
distinction.  Even so systematic variations of any kind from the Salpeter
slope have not been measured outside of some evidence for non-standard IMFs in
low surface brightness galaxies (LSB) and galaxies experiencing powerful
bursts of star formation \citep{Elmegreen06}.  Either way, observational
evidence for systematic variations of the IMF in galaxies would be highly
valuable. 

The plan of this paper is as follows:  \S 2 explains our method for
constraining the IMF.  In \S 3 we describe the SDSS data and our sample
selection.  \S 4 details our modeling scheme.  In \S 5 we discuss our
statistical techniques.  \S 6 reports our results.  In \S 7 we check our
results against the H$\delta_{\rm A}$ distribution of the data and \S
8 presents our conclusions.

\section{Methodology}

This paper revisits the ``classic'' method of K83 (and the subsequent
extension KTC94) to constrain the IMF of integrated stellar populations.  The
method takes advantage of the sensitivity of the H$\alpha$ equivalent width
(EW) to the IMF.  K83 showed that model IMF tracks can be differentiated in
the $(B-V) - \log{(\rm{H}\alpha\rm{\ EW})}$ plane.

The total flux of a galaxy at 6565\AA\ is the combination of the
underlying continuum flux plus the flux contained in the H$\alpha$
emission line.  The H$\alpha$ flux and the continuum flux have
different physical origins, both of which can be used to gain physical
insights into galaxies.

In the absence of AGN activity the H$\alpha$ flux is predominantly
caused by massive O and B stars which emit ionizing photons in the
ultraviolet.  O and B stars are young and found in the regions of
neutral hydrogen in which they formed.  In Case B recombination, where
it is assumed that these clouds are optically thick, any emitted Lyman
photons are immediately reabsorbed.  After several scattering events
the Lyman photons are converted into lower series photons  (including
H$\alpha$) and two photon emission in the $2^2 S \rightarrow 1^2 S$
continuum.  These photons experience smaller optical depths and can
escape the cloud.  The transition probabilities are weakly dependent
on electron density and temperature and can be calculated.  Through
this process the measured H$\alpha$ flux can be converted into the
number of O and B stars currently burning in an integrated stellar population. 

However Case B recombination is an idealized condition and it is possible that
ionizing photons can escape the cloud without this processing, a situation
known as Lyman leakage.  As such the H$\alpha$ flux is a lower limit on the
number of O and B stars present.

The continuum flux of a galaxy is due to the underlying stellar
population.  At 6565\AA\ the continuum is dominated by red giant stars
in the 0.7-3 M$_\sun$ range while the H$\alpha$ flux comes from stars more
massive than 10 M$_\sun$.  

The EW is defined as the width in angstroms of an imaginary box with a height
equal to the continuum flux level surrounding an emission or
absorption line which contains an area equal to the area contained in
the line.  This is effectively the ratio of the strength of a emission
or absorption line to the strength of the continuum at the same
wavelength.  Given the physical origins of the H$\alpha$ flux and the
continuum at 6565\AA\ the H$\alpha$ EW is the ratio of massive O and B
stars to stars around a solar mass.  Therefore the H$\alpha$ EW is
sensitive to the IMF slope above around 1 M$\sun$ and can be used to
probe the IMF in galaxies.

As mentioned in the introduction several degeneracies plague the study of the
IMF from the integrated light properties of galaxies.  Variations in the IMF,
age, metallicity, and SFH of galaxy models can all yield similar effects in the
resulting spectra.  For example, increasing the fraction of massive stars,
reducing the age of a galaxy, lowering the metallicity, and a recent increase
in the star formation rate will all make a galaxy bluer.

Metallicity effects were not discussed in either K83 or KTC94 and galaxy ages
were assumed, 15 Gyr for K83 and 10 Gyr for KTC94.  

The SFH in K83 is addressed by calculating models with exponentially decreasing
SFHs for a range of e-folding times, as well as a constant and a linearly
increasing SFH.  In the $(B-V) - \log{(\rm{H}\alpha\rm{\ EW})}$ plane the
effect of varying the SFH e-folding time is orthogonal to IMF variations.
However this is only true for smoothly varying exponential and linear SFHs.
Discontinuities, either increases (bursts) or decreases (gasps), in the star
formation rate can affect the H$\alpha$ EW relative to the color in ways
similar to a change in the IMF.  Along with the exponential SFHs KTC94 also
uses models with instantaneous bursts on top of constant SFHs.  However this
was done to access high EWs at an age of 10 Gyr rather than to fully flesh out
the effects of SFH discontinuities. 

The assumption of smoothly varying SFHs is a key assumption in our analysis.
Most late-type galaxies are thought to form stars at a fairly steady rate over
much of recent time although bursts of star formation may play a significant
role in low mass galaxies \citep{Ken98b}.  For these galaxies smoothly varying
SFHs are justified.  However there are other galaxies clearly in the midst of a
strong burst of star formation (e.g. M82) and dwarf galaxies with complex SFHs,
e.g. NGC 1569 \citep{Angeretti05}, for which this assumption is a poor one.
The effects of violations of our SFH assumptions are described in detail in the
results section.

\section{The Data}

K83 cites four major sources of error all of which are improved upon or
eliminated by the high, uniform quality of SDSS spectroscopic and photometric
data.  

The H$\alpha$ fluxes in his sample can be contaminated by nonthermal
nuclear emission.  In the updated investigation, KTC94, this problem is
addressed by removing objects with known Seyfert or LINER activity and
luminous AGN.  The SDSS spectra allow measurements of emission line ratios
which can be used to separate star forming galaxies from AGN \citep{BPT}.

The second problem is that the H$\alpha$ emission flux will be underestimated
if the underlying stellar absorption of H$\alpha$ is not taken into
consideration.  The narrow band filter photometry of \citet{KK83} could not
measure this effect for individual galaxies so a fixed ratio was assumed for
all galaxies.  The SDSS spectroscopic pipeline does not take this into account
either. We use the H$\alpha$ fluxes measured from the SDSS spectra by
\cite{CAT} which fit the continua with stellar population models to more
accurately measure the H$\alpha$ emission.  While the SDSS pipeline method is
sufficient for strong emission lines the H$\alpha$ absorption EW can be as
large as 5 \AA\ which is significant for weaker emission lines.

Thirdly, their narrow band H$\alpha$ imaging includes [\ion{N}{2}] emission
which is corrected for by assuming a constant [\ion{N}{2}]/H$\alpha$ ratio.
The H$\alpha$ and [\ion{N}{2}] emission lines are resolved in the SDSS spectra
so there is no need for a correction.  This is a significant improvement.
\citet{KK83} found from a literature survey that the mean value of the
H$\alpha$/(H$\alpha$ + [\ion{N}{2}]) ratio is $0.75 \pm 0.12$ for spiral
galaxies and $0.93 \pm 0.05$ in irregular galaxies.  These mean corrections
were applied uniformly to the K83 data.  In KTC94 a uniform correction was
applied using [\ion{N}{2}]/H$\alpha$ = 0.5. For comparison in our sample the
mean value of the H$\alpha$/(H$\alpha$ + [\ion{N}{2}]) ratio is a strikingly
similar 0.752 and the mean [\ion{N}{2}]/H$\alpha$ ratio is 0.340.  However
[\ion{N}{2}]/H$\alpha$ ranges from 0.0 to 0.6.  If our H$\alpha$ and
[\ion{N}{2}] lines were blended, applying a fixed correction would introduce
errors of as much as 25\% in the H$\alpha$ EWs of individual galaxies. 

Lastly, in both K83 and KTC94 extinction corrections were addressed by
plotting data alongside models which were either assumed an average value for
the extinction or were unextincted.  The Balmer decrement (H$\alpha$/H$\beta$)
can be measured from SDSS spectra which allows for extinction corrections for
individual galaxies.

Another major advantage of this study is that the sample size is much larger
than that of K83 and KTC94.  The KTC94 sample contains 210 galaxies, whereas
ours has $\sim 10^5$.  The large sample size allows us to investigate IMF
trends as functions of galaxy luminosity, redshift and aperture fraction with
subsamples larger than the entire KTC94 sample.

There is a key disadvantage to this method as well.  K83 and KTC94 were able
to adjust the sizes of their photometric apertures to contain the entire disk
of individual galaxies to a limiting isophote of 25$\mu$ given by the RC2
catalog \citep{KK83}.  The advantage of narrow band measurements of H$\alpha$
EW is that they can cover a much larger aperture and match the broadband
measurements set to match the physical size of individual galaxies.  The SDSS
has fixed 3" spectroscopic apertures.  This problem is partly offset by using
matching 3" photometry apertures from the SDSS.  However this introduces
aperture effects as observed galaxies have a wide range of angular sizes due
to the range of physical sizes and distances present in the local universe.
This is significant as radial metallicity gradients (e.g. \citet{VC92}) are
observed in spiral galaxies which could incorrectly be interpreted as radial IMF gradients.

Even so, this method can constrain the IMF within the SDSS apertures.  For our
program galaxies 23\% of the total light falls in the SDSS apertures.  The
fact that the more distant galaxies are more luminous and tend to be larger
helps to balance out the larger physical scales of the fixed aperture size at
greater distances.  On average 17\% of the light falls in the aperture for the
faintest galaxies while it is 25\% for the brightest bin.  In spite of their
limited size the SDSS apertures still contain a great diversity of stellar
populations which make this data set an excellent test bed for IMF
universality.  We will present extensive tests of aperture effects below.

\subsection{Sample Selection}

The sample is selected from Sloan Digital Sky Survey data.  The project goal
of the SDSS is to image a quarter of the sky in five optical bands with a
dedicated 2.5 m telescope \citep{York00}.   From the imaging $10^6$ galaxies
and $10^5$ quasars will be selected for spectroscopic followup.  Photometry is
done in the $ugriz$ filter system described by \citet{F96}.  Magnitudes are on
the \texttt{arcsinh} system \citep{LGS99} which approaches the AB system with
increasing brightness.  Spectra are taken with a multi-object fiber
spectrograph with wavelength coverage from 3800\AA\ to 9200\AA\ and $R \sim
1800$ \citep{U99}.

Our sample is a sub-sample of the Main Galaxy Sample from SDSS DR4
\citep{DR4}.  The Main Galaxy Sample (MGS) targets galaxies with $r \leq
17.77$ in Petrosian magnitudes \citep{EDR}.  All galaxies in the MGS are
strong detections so the differences between luptitudes and the AB system can
be ignored.  In order to avoid fiber crosstalk in the camera an upper
brightness limit of $g=15.0$, $r=15.0$ and $i=14.5$ is imposed.  Targets are
selected as galaxies from the imaging by comparing their PSF magnitudes to
their de Vaccouleur's and exponential profile magnitudes.  Exposure times for
spectroscopy are set so that the cumulative median signal-to-noise satisfies
$(\rm{S/N})^2 > 15$ at $g=20.2$ and $i=19.9$ fiber magnitudes.  The time to
achieve this depends on observing conditions but always involves at minimum
three 15 minute exposures.

Due to the construction of the spectrograph fibers cannot be placed closer
than 55" to each other.  This may be a source of bias in the sample.  Cluster
galaxies may be preferentially excluded from the sample.  The SDSS
collaboration has plans to quantify this effect in the near future
\citep{DR4}.  LSBs are excluded from the MGS with a surface brightness cut
which may also bias the sample \citep{EDR}. There is some evidence that LSBs
may have IMFs which differ from a universal IMF.  \citet{Lee04} find that the
comparatively high mass-to-light ratios of LSBs can be explained with an IMF
deficient in massive stars relative to normal galaxies.

Our sample begins with the fourth data release (DR4) of the SDSS \citep{DR4}.
DR4 covers 4783 deg$^2$ in spectroscopy for a total of 673,280 spectra,
567,486 of which are galaxy spectra.  429,748 of these have flags set
indicating they are part of the MGS.  DR4 also includes special spectroscopic
observations of the Southern Stripe which were not selected by the standard
algorithm but which nonetheless have the \texttt{TARGET\_GALAXY} flag set in
\texttt{primTarget} which usually indicates membership in the MGS.  These
objects are identified by comparing their spectroscopic plate number to the
list of special plates in \citep{DR4} and rejected.  This leaves 423,285
spectra.

The first round of cuts to our sample address general data quality.  While the
overall quality of SDSS data is high there are a handful of objects with
pathological values for one or more parameters.  First we require that all
parameters of interest have reasonable, real values.  This means that the
Petrosian and fiber magnitudes must be between 0 and 25 and have errors smaller
than 0.5 in all five $ugriz$ bands.  Line flux errors are capped at $10^{-12}
{\rm \ erg\ cm^{-2}\ s^{-1}}$ \AA$^{-1}$ and equivalent widths at $10^5$\AA.
For most of the parameters less than 1\% of objects fail this loose
requirement.  However 6.6\% of the objects fail the Petrosian magnitude error
requirement.  This is most likely due to the photometric pipeline having a
difficult time defining the Petrosian radius.  As such this constraint is
potentially biased against LSBs or galaxies with unusual morphologies.  In
defense of this cut we later bin our data by luminosity and aperture fraction
both of which are determined in part by the Petrosian magnitudes and also by
$K$-corrections determined from them.  In addition, limiting flux errors to
50\% is hardly unreasonable.  Altogether 391,160 galaxies pass these combined
requirements, which is 92.4\% of the MGS.

Galaxies from photometry Run 1659 are removed because of a known problem with
the photometry.  This excludes 4,485 galaxies (1.1\%) from a continuous strip
on the sky and should not be a source of bias.  We place a further constraint
on the $z$ band fiber magnitude requiring that the error be less than 0.15.
The $z$ band generally suffers from the most noise so this requirement ensures
that the fiber magnitude quality is good enough to minimize the chance of
erroneous $K$-corrections which could affect our colors.  Only 2,866 (0.7\%) MGS galaxies fail this test.

Combining the general data quality requirements, the Run 1659 rejection and
the fiber $z$ band error limit leaves 386,647 galaxies (91.3\% of the MGS).

The next round of cuts to our sample, while necessary, have clear
astrophysical implications.  Many of the objects in the MGS have AGN components.  As we are interested in studying only the underlying stellar populations of these objects AGN must be removed.  This is done using the classical \citet{BPT} diagram comparing the logarithms of the [\ion{O}{3}
$\lambda 5007$]/H$\beta$ and [\ion{N}{2} $\lambda 6584$]/H$\alpha$ emission
line ratios.  We used the criterion of \citet{Kauf03} where objects for which
\begin{equation}
\log([\rm{O\ III}]/\rm H\beta) > \frac{0.61}{\log([\rm{N\ II}]/H\alpha)-0.05}+1.3
\label{eq:KBPT}
\end{equation}
are classified as AGN and rejected.   Following \citet{Brinch04} we require
the S/N of the H$\alpha$, H$\beta$, [\ion{O}{3}] and [\ion{N}{2}] lines to be
at least 3 to properly classify a galaxy as a star forming one.  131,807
galaxies (34.1\% of MGS objects surviving our first round of cuts) survive
this cut.

The above cut automatically rejects any galaxies with weak [\ion{O}{3}] and
[\ion{N}{2}] lines.  This excludes galaxies with weak star formation.  To a
lesser extent metal poor galaxies which also have weak [\ion{N}{2}] emission
are rejected as well.  \citet{Brinch04} also define a low S/N star forming
class of galaxies which we identify and keep in our sample.  These are the
galaxies which have not already been classified as star forming or AGN by
strong lines and equation \ref{eq:KBPT}, nor have been identified as low S/N
AGN by [\ion{N}{2} $\lambda 6584$]/H$\alpha$ $>$ 0.6 with S/N $>$ 3 in both
lines, yet still have H$\alpha$ with S/N at least 2.  79,548 (20.6\%) of the
sample falls into the low S/N star forming galaxy category.  Combining the two
classes 211,355 (54.7\%) of the galaxies survive the AGN cut. 

The AGN cut also has a strong luminosity bias for two reasons.  Galaxies
with AGN components tend to be brighter.  The bimodal distribution of galaxies
in color-magnitude space \citep{Baldry04} also plays a role.  The most
luminous galaxies are predominantly red with minimal star formation and thus
weak emission lines.  Luminous galaxies are rejected both for having AGN
components and for having low S/N emission lines.  Over 95\% of galaxies
fainter than ${\rm M}_{r,0.1} = -19$ meet this criteria, but by ${\rm
  M}_{r,0.1} = -24$ the fraction is only 38.9\%.

A color bias is also introduced by the AGN cut.  Over 95\% of galaxies bluer
than $(g-r)_{0.1} = 0.6$ pass, which drops to 24\% by $(g-r)_{0.1} = 1.2$.
This is mainly due to the S/N requirement for the emission lines.  Redder
galaxies tend to have weak emission lines and are rejected. 

The Balmer decrement is used for the extinction correction so H$\alpha$ and
H$\beta$ S/N are required to be at least 5 to reduce errors.  214,912 galaxies
(55.6\%) have H$\beta$ S/N $> 5$ which is the more restrictive of the two
criteria.  This cut has a clear luminosity bias.  Roughly 85\% of galaxies
with ${\rm M}_{r,0.1} > -20$ satisfy this requirement, but this fraction
decreases with increasing luminosity until only 12.9\% survive at ${\rm
  M}_{r,0.1}=-24$.  This is again due to the bimodal distribution of galaxies.
The luminous red galaxies with weak emission lines are rejected.

There is also a color bias.  Over 93\% of the bluest galaxies blueward of
$(g-r)_{0.1}=0.8$ survive the cut while only 23\%  of the reddest pass this
requirement.  As previously mentioned, by nature the reddest galaxies have
weak Balmer lines as a result of their low SFRs and are preferentially
rejected.

A redshift cut of $0.005 \leq z \leq 0.25$ is applied to ensure that peculiar
velocities do not dominate at low redshift and to limit the range of galaxy
ages.  384,349 galaxies (99.4\%) meet this criteria.  Nearly all galaxies from
${\rm M}_{r,0.1}=-17$ to $-23$ survive this cut.  On the low luminosity end
only 27.4\% of ${\rm M}_{r,0.1}=-14$ galaxies are distant enough to pass and
on the high end 93.1\% of ${\rm M}_{r,0.1}=-24$ galaxies are close enough to
survive.  Only 63\% of the bluest galaxies pass.  Many of the blue galaxies
that fail are actually \ion{H}{2} regions of Local Group galaxies which are
treated as their own objects by the SDSS pipeline so removing them actually
improves the integrity of our sample.

The stellar populations of galactic bulges can be significantly different from
those in the spiral arms.  The SDSS fibers have a fixed aperture of 3'' so
over the large range of luminosities and distances in the MGS aperture affects
can become very important.  To remove outliers we require that at least 10\%
of the light from a galaxy falls within the spectroscopic aperture.  This is
done by comparing the Petrosian magnitude to a fixed 3'' aperture fiber
magnitude.  Both of these quantities are calculated for all objects in the
SDSS by the photometric pipeline.  371,777 (96.2\%) galaxies survive this cut.
This cut rejects proportionally more faint galaxies; 99.2\% pass at
${\rm M}_{r,0.1}=-24$ compared to 50.9\% at ${\rm M}_{r,0.1}=-14$.  The
aperture cut has low sensitivity to color.

The intersection of the AGN, Balmer line S/N, redshift and aperture fraction
cuts leaves 140,598 galaxies-- 36.4\% of the high quality MGS data defined by
our first round of cuts and 33.2\% of the MGS as a whole.

At this point three final cuts are applied to the sample.  One galaxy is
removed for surviving all criteria, but having a negative H$\alpha$ EW.  

The extinction of individual galaxies is estimated using the Balmer decrement
(H$\alpha$/H$\beta$ emission flux ratio) for each galaxy.  Given Case B
recombination, a gas temperature of 10,000K and density of 100 $\rm{cm}^{-3}$
the Balmer decrement is predicted to be 2.86 \citep{O89}.  This ratio is
weakly dependent on nebular temperature and density.  \citet{O89} lists values
down to 2.74 for Case B recombination in environments where both the
temperature and electron density are high.  3.2\% of the galaxies suffer from
the problem that the Balmer decrement is less than 2.86 and 2.1\% have a
Balmer decrement below 2.74.  538 galaxies (0.4\%) have Balmer decrements more
than 3$\sigma$ below 2.74, which is around 6 times more than expected.  This
is does not suggest a problem with the Case B assumption as the predicted
Balmer decrements for Case A recombination are nearly identical in each
temperature regime.

To understand the reason behind this problem around 100 of the offending spectra
were inspected revealing a few different causes for the problem.  Around 80
galaxies are at redshifts where the telluric \ion{O}{1} $\lambda 5577$ line
affects the measurement of H$\beta$.  Many of these galaxies have very strong
emission lines with extremely weak stellar absorption.  Using the SDSS
pipeline values instead of the \citet{CAT} values yields acceptable Balmer
decrements.  The rest are galaxies with low flux where the Balmer lines are in
absorption.  This shows there are a few cases where attempting to fit the
underlying stellar absorption lines fails and this failure is not reflected in
the error values.  These galaxies are rejected without any apparent
introduction of bias.

All of these cuts combined leave 140,060 galaxies, which is 33\% of the MGS
and 36\% of the high quality MGS data.  After removing duplicate observations
there are 130,602 galaxies in our sample.  Of these objects 1.7\% overlap with
the Luminous Red Galaxy Sample.

Overall the bulk of the galaxies are removed by AGN rejection and the H$\beta$
S/N requirement, with the rest of the cuts having little effect.  Both of
these cuts are necessary.  AGN must be removed to ensure that the H$\alpha$
emission represents the underlying stellar population and not an accretion
disk.  Our method requires that the galaxies have measurable Balmer emission
lines.  This coupled with the need for accurate extinction corrections
justifies the Balmer line S/N requirement.

Our cuts bias our sample by preferentially excluding galaxies at both
luminosity and both color extremes.  The faintest galaxies are most affected
by the redshift and aperture cuts while the luminous galaxies succumb to the
H$\beta$ S/N requirement.  At the red extremes it is the H$\beta$ S/N and AGN
requirements that play equally large roles, while the bluest objects are
primarily rejected by the Hubble flow redshift requirement.  Although our
sample is biased by our cuts each one is a necessary evil.  We do not attempt
to correct this bias, but we remind the reader that the following results are
only representative of actively star forming galaxies without any AGN activity.

The aim of this paper, however, is to test IMF \textit{universality}.  If the
IMF is truly universal it should be universal in any subsample of galaxies.
The fact that our sample is slightly biased with respect to luminosity and
color is not a significant barrier to achieving our goal.

\subsection{Corrections}

The SDSS includes a number of different calculated magnitudes.  We use the
\textit{fiber} magnitudes which are 3'' fixed aperture magnitudes.  Although
it was not the case in earlier versions of the photometric pipeline, fiber
magnitudes are now seeing corrected \citep{DR2}.  The fiber magnitudes were
not originally intended for science purposes but rather to get an idea of how
bright an object will appear in the spectrograph.  We use the fiber magnitudes
to reduce the aperture effects arising from comparing a 3'' spectroscopic
aperture to Petrosian magnitudes.  Originally the SDSS used ``smear''
exposures to correct spectra for light falling outside the 3'' aperture due
seeing, guiding errors and atmospheric refraction \citep{EDR}.  The smear
technique was later found to be an improvement only for high S/N point sources
and its use was discontinued \citep{DR2}.  The spectra here are not seeing
corrected. 

After paring the sample to its final size a number of corrections must be made
to both the photometric and spectroscopic data.  Galactic reddening from the
Milky Way must be corrected for.  SDSS database includes the \citet{SFD} dust
map values for each photometry object.

The extinction of individual galaxies is estimated using the Balmer decrement
for each galaxy.  The data is corrected assuming that 2.86 is the true value
of the Balmer decrement using the Milky Way dust models of \citet{Pei92}.  The
assumption of Milky Way dust is not significant as models of the dust
attenuation in the Milky Way, SMC and LMC are nearly identical in the $g$ band
and redward.  As aforementioned a few percent of our galaxies have Balmer
decrements below 2.86.  Our solution is to set the emission line extinction to
${\rm A}_{{\rm V},l} = 0.01$ magnitudes for these galaxies.

Massive young stars and their surrounding ionized nebulae tend to be embedded
in their star forming regions more so than older, lower mass stars which have
had time to migrate from their birth regions.  As such nebular emission lines
will experience more extinction than the continuum.  \citet{Calzetti94} find
the ratio of emission to continuum line extinction is $f = 2.0 \pm 0.4$.  We
assume this value to be 2.0 and correct the continuum and emission lines
separately.  This is the same value used by K83 and KTC94 in their extinction
corrected models.  We note that the spatial geometry of the dust can influence
the extinction law, but this complication is beyond the scope of this paper.

Galaxy photometry is $K$-corrected to $z=0.10$ using version 4.1.4 of the code
of \citet{Blanton03}.  This redshift is roughly the median of the sample and
is selected to minimize errors introduced by the $K$-corrections.  The
$(g-r)_{0.1}$ colors we use are the $g-r$ colors we would observe if the
galaxies were all located at $z=0.1$.

Stated explicitly the $(g-r)_{0.1}$ color is
\begin{equation}
c=(g-r)_{0.1} = (g - k_g - {\rm A}_g - 1.153  {\rm A}_{{\rm V},l}/ f) -
(r - k_r - {\rm A}_r - 0.834  {\rm A}_{{\rm V},l}/ f)
\label{eq:color}
\end{equation}
where $k_g$ and $k_r$ are $K$-corrections, ${\rm A}_g$ and ${\rm A}_r$ are
Milky Way reddening values, and 1.153 and 0.834 relate the V band extinction
to the $g$ and $r$ bands assuming a Milky Way dust model.  The corrected
equivalent width is obtained as follows
\begin{equation}
w = w_0 \left [ (1+z) \times 10^{-0.4(0.775 {\rm A}_{{\rm V},l}) (1-1/f)} \right ]^{-1}
\label{eq:width}
\end{equation}
where $w_0$ is the measured, uncorrected equivalent width.  The $1+z$ arises
from the fact that the total flux in the H$\alpha$ line is not affected by
cosmological expansion of the universe but the flux per unit wavelength of the
continuum is depressed by a factor of $1+z$.  The following term is the
extinction correction.  The 0.775 relates the V band extinction to the
H$\alpha$ line assuming a Milky Way dust model and the $1-1/f$ is due to the
fact that the emission line and continuum experience different amounts of
extinction as previously explained.

\begin{figure}
\epsscale{0.6}
\plotone{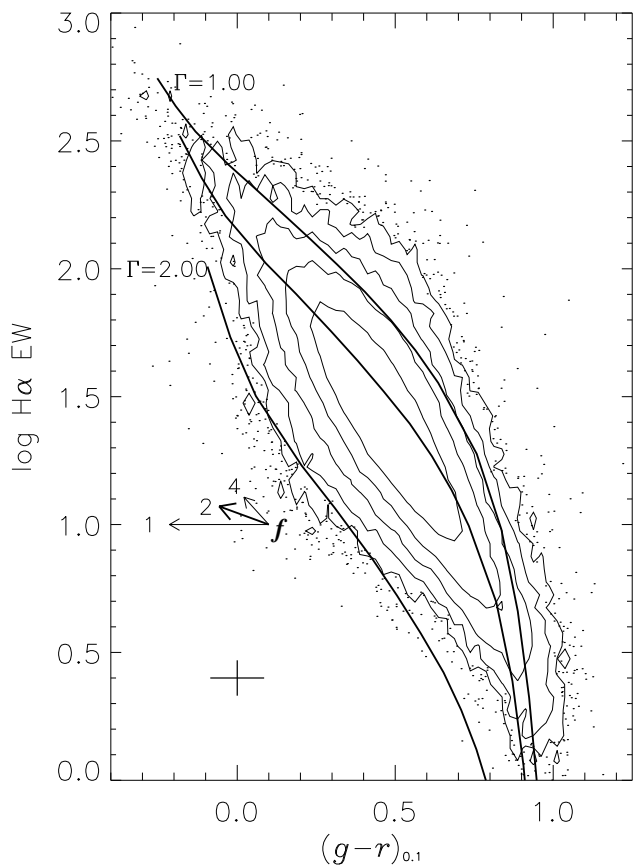}
\caption{Distribution of all 130,602 program galaxies in the $(g-r)_{0.1}$ vs. $\log(\rm{H}\alpha \rm{\ EW})$ plane.  The contours are logarithmic.  Outside the last contour individual points are plotted.  The thick lines are model tracks with exponentially decreasing SFHs with $\tau = 1.1$ Gyr and solar metallicity.  The age increases along the tracks from 100 Myr in the upper left to 13 Gyr in the lower right.  The upper line has $\Gamma=1.00$, the middle line is similar to Salpeter's IMF with $\Gamma=1.35$ and the lower line has $\Gamma=2.00$.  They are identical to the solid lines in Figure \ref{fig:models}.  The cross in the lower left indicates the median error bars of the sample.  The arrows are dust vectors are for typical observed Balmer decrements (H$\alpha$/H$\beta$ = 4) for different values of $f$.  The values of $f$ are 1, 2 and 4 clockwise from the one pointing left. }
\label{fig:grha_data}
\end{figure}

Figure \ref{fig:grha_data} shows the distribution of the galaxies in the color
vs. H$\alpha$ EW plane.

\subsection{Errors}

In order to conduct a likelihood analysis we need error estimates which take
into account both the errors induced by the photometry and spectroscopy and
those by the aforementioned corrections.  The error in the corrected color,
$\sigma_c$, is given by 
\begin{equation}
\sigma_c = 0.03 + \sqrt{\sigma_g^2 + \sigma_r^2 + \left (\frac{0.319 {\rm A}_{{\rm V},l}}
{f}\right )^2 \left ( \left ( \frac{\sigma_{{\rm A}_{{\rm V},l}}}{{\rm A}_{{\rm V},l}} \right )^2 + 
\left (\frac{\sigma_f}{f}\right )^2\right ) + \left(0.0440{\rm A}_g \right)^2
+ \sigma_k^2}
\label{eq:sigmac}
\end{equation}
where $\sigma_g$ and $\sigma_r$ are the Poisson errors in the observed $g$ and
$r$ band photometry, $f$ is the ratio of the emission line to continuum
extinction, and $\sigma_k$ is the error introduced by the $K$-corrections.
The terms inside the square root in equation \ref{eq:sigmac} are obtained
through a straight error propagation of equation \ref{eq:color}.  The 0.03
outside the square root is due to the systematic zero point errors of the SDSS
filter system.  Following \citet{Calzetti94} $f$ is fixed at 2.0 and
$\sigma_f$ is set to 0.4. The error in emission line $\rm{A_V}$ is given by
\begin{equation}
\sigma_{{\rm A}_{{\rm V},l}}^2 = 9.440\left( \left( \frac{\sigma_{\rm{H}\alpha}}
{\rm{H}\alpha} \right)^2 +  \left(\frac{\sigma_{\rm{H}\beta}}
{\rm{H}\beta} \right)^2 \right)
\label{eq:sigmaav}
\end{equation}
which is dependent on the fractional uncertainty of the H$\alpha$ and H$\beta$
emission line fluxes.  \citet{SFD} reports a 16\% error in their Milky Way
reddening values.  Because a dust model is assumed reddening in the $g$ and
$r$ bands are linearly related, so the error in ${\rm A}_g - {\rm A}_r$ is a
function of ${\rm A}_g$.  This relationship combined with the 16\% error
yields the 0.0440 in equation \ref{eq:sigmac}.  The median value of $g$ band
reddening is 0.10 so the errors introduced by the MW reddening correction are
insignificant for the majority of objects.  Errors in the redshift
determination are negligible, typically 0.01\%.  The value of $\sigma_k$ is estimated to be
0.02 by visual inspection of a plot of $k_g-k_r$ as a function of redshift. 

For a typical galaxy the term involving $\sigma_f$ is the largest contributor
to the extinction corrected color error.  The median Poisson error from the
photometry is 0.01 in both bands.  The median value of $\sigma_c$ is 0.085.

The error in the corrected H$\alpha$ equivalent width, $\sigma_w$, is given by
\begin{equation}
\sigma_w^2 = w^2 \left[ \left( \frac{\sigma_{w_0}}{w_0}\right )^2 +
0.5095\left( \left( \sigma_{{\rm A}_{{\rm V},l}}(1-1/f) \right)^2 +
\left( \frac{\sigma_f {{\rm A}_{{\rm V},l}}}{f^2} \right)^2 \right) \right]
\label{eq:sigmaw}
\end{equation}
Equation \ref{eq:sigmaw} is the result of propagating the errors in equation
\ref{eq:width}, neglecting the insignificant redshift errors.  Again, the term
involving $\sigma_f$ is the largest contributor to the error for typical
galaxies.  The median error in the equivalent width is 17\%.  The median error
bars for the sample are shown in Figure \ref{fig:grha_data}.  For comparison,
K83 reports equivalent width errors of around 10\%, but the extinction is
uncertain at the 20-30\% level.

\section{Models}

Model galaxy spectra were calculated using the publicly available PEGASE.2
spectral synthesis code \citep{Fioc97}.  Models are calculated for ages from 1
Myr to 13 Gyr.    25 smoothly varying SFHs generated by analytic formulae are
considered.  The SFHs range from 19 exponentially decaying SFHs with time
constants from 1.1 to 35 Gyr, a constant SFR, and four increasing SFHs which
are proportional to $1-\exp^{-t/\tau}$ where $\tau$ is the time constant.  The
precise values of the time constants were selected to smoothly sample the
H$\alpha$ EW vs. $(g-r)_{0.1}$ plane.  The metallicity of the stars is assumed
to be constant with respect to time and calculated for $Z$ = 0.005, 0.010,
0.020 and 0.025.  Galactic winds, galactic infall and dust extinction are
turned off.  The dust extinction is not modeled because we have applied an
extinction correction to the data.  Nebular emission is calculated from the
strength of the Lyman continuum.  Emission line ratios are fixed.  The model
spectra are redshifted to $z=0.1$ to match the redshift range of the data.

The model parameter of interest is the IMF. Implicit in equation
\ref{eq:imfbaldry} is the assumption that the IMF does not vary as a function
of time.

The continua of galaxies are weakly influenced by low mass stars in the
optical.  This method is sensitive to the IMF for masses above around 1
M$_\sun$ so the slope is fixed below 0.5 M$_\sun$.  Above 0.5 M$_\sun$ models
are calculated for $1.00 \leq \Gamma \leq 2.00$, where $\Gamma$ is incremented
by 0.05 between models. 

We treat our IMF model as though it has only one degree of freedom-- $\Gamma$
above 0.5 M$_\sun$.  In truth it has three more as written: the lower and
upper mass cutoffs and the point at which the slope changes.  So how well are
our assumptions justified?

We have parameterized the IMF as a piecewise power law with two components.
Piecewise power laws are motivated by empirical fits to data starting with
\citet{S55}, which had only one component.  By contrast the power law
formulation of the \citet{Scalo86} IMF has 24 components.  The log-normal
distribution is a more physical choice as it can arise from stochastic
processes.  \citet{MS79} were the first to fit an observational measurement of
the IMF with a log-normal distribution.  The log-normal distribution is
normalizable as it goes to zero smoothly at both extremes without any awkward
truncation.  Its main drawback is that it cannot fit any structure in the IMF
over small mass ranges.

Log-normal distributions have three degrees of freedom.  This is less than the
four our model has.  However, we are not sensitive to IMF over the full range
of masses which makes it much more difficult to fit the parameters of the
log-normal distribution.  Instead we use this piecewise model and lower the
degrees of freedom through physical arguments.  Many investigators find a
change in slope in the IMF around 0.5 M$_\sun$.  Our fixed lower end of the
IMF is designed to be consistent to this.  As our technique is not sensitive
to this regime this assumption does not impact the results.

The IMF must be normalizable because the total mass of a stellar population is
finite.  In our parameterization this is achieved by truncation at 0.1 and 120
M$_\sun$.  This seems unphysical as the existence of brown dwarfs suggests
that the IMF should continue below the hydrogen burning limit.  However, stars
at 0.1 M$_\sun$ do not contribute much to the integrated light of galaxies.
As we are not sensitive to stars in this mass range this choice is not
unreasonable.  In fact, truncating the IMF at 0.5 M$_\sun$ yields models which
are at worse differ by 0.002 in $(g-r)_{0.1}$ and 3\% in H$\alpha$ EW from
those with low mass stars.  The slope below 0.5 M$_\sun$ has essentially no
effect on our results.  Only when the IMF is truncated at 0.9 M$_\sun$ do the
models differ at the level of the errors in the data. 

On the high mass end the choice of limit does matter.  There is a physical
upper limit to the size of stars associated with the Eddington limit.  The
value of this theoretical limit is not widely agreed upon.  The largest
stellar mass measured reliably, via analysis of a binary system, is $83 \pm 5$
M$_\sun$ \citep{Bon04}.  \citet{WK04} argue that given the large mass and
youth of the star forming cluster R136 in the Large Magellanic Cloud stars
in excess of 750 M$_\sun$ should be present given a Salpeter IMF with no upper
mass limit to stars, whereas no stars above 150 M$_\sun$ are observed.  An
analysis of the Arches Cluster, the youngest observable cluster, gives an
upper limit of 150 M$_\sun$ based on Monte Carlo simulations although stars
above 130 M$_\sun$ are not detected \citep{Figer05}.  The PEGASE model tracks
only extend up to 120 M$_\sun$ so this is the cutoff used.  Another issue is
that the physics and evolution of such high mass stars is not well known so
the models themselves may be a significant source of error in this regime.

The left half of Figure \ref{fig:oddimf} shows the effects of varying the high
mass cutoff in the IMF.  The effect in the $(g-r)_{0.1}$-H$\alpha$ EW plane is
seen to be very similar to increasing the value of $\Gamma$.  Lowering the
upper mass cutoff from 120 to 90 M$_\sun$ has roughly the same effect as
increasing $\Gamma$ from 1.35 to 1.45.  The relationship between the change in
the upper mass cutoff (from 120 M$_\sun$) and the apparent change in $\Gamma$
is $\Delta \Gamma \sim 0.005 \Delta {\rm M_{up}}$ and is roughly linear for
upper mass cutoffs down to 50 M$_\sun$.  The coefficient in the relationship
is a week function of the age of the population ranging from 0.004 for 13 Gyr
old populations to 0.006 for 100 Myr old populations.

The right half of Figure \ref{fig:oddimf} shows the affect of adding a second
break in the IMF at 10 M$_\sun$.  Reducing the value of $\Gamma$ over the 0.5-10
M$_\sun$ range while keeping it fixed at $\Gamma=1.35$ above 10 M$_\sun$ has a
similar effect to decreasing $\Gamma$ in a two component model.  This
illustrates one of the limitations of this model.  At this point it is not
possible to detect fine structure in the IMF slope or to state precise values
for the IMF slope.  In this limited space of observables the IMF models
themselves are degenerate.  What it does provide is a framework with which to
detect variations in the IMF.  Although we can construct similar tracks from
different IMF models, we can still detect the differences between two groups
of galaxies.

While we will report our results as a function of $\Gamma$ it must always be
kept in mind that it is degenerate with the upper mass cutoff and other fine
structure in the IMF at high stellar masses.

The assumption of smoothly varying SFHs is of great consequence.  In the event
that a galaxy is experiencing or has recently experienced a burst our SFH
assumption can lead to measured $\Gamma$ values that are off by as much as
0.5.  The effects of bursts are more closely examined in a later section.

\begin{figure}
\epsscale{0.8}
\plotone{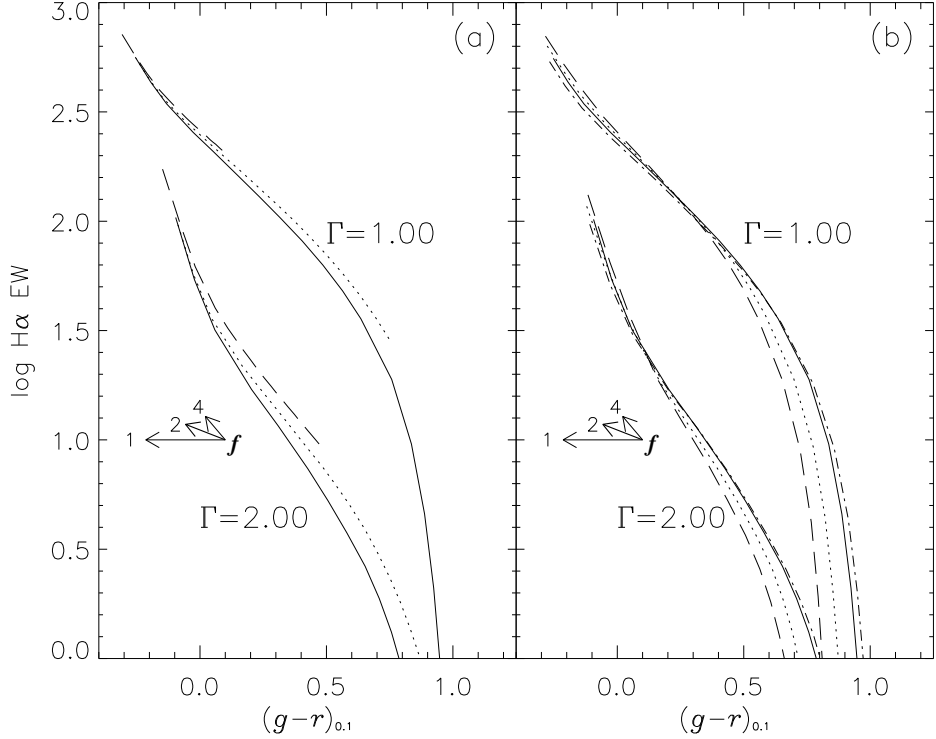}
\caption{Effects of model parameters in the $(g-r)_{0.1}$ vs. $\log(\rm{H}\alpha \rm{\ EW})$ plane.  In both panels ages of the stellar population increase along the tracks from 100 Myr on the upper left to 13 Gyr on the lower right.  Also in both panels the upper set of tracks have $\Gamma=1.00$ while the lower set have $\Gamma=2.00$.  The arrows are dust vectors are for typical observed Balmer decrements (H$\alpha$/H$\beta$ = 4) for different values of $f$.  The values of $f$ are 1, 2 and 4 clockwise from the one pointing left.  (a) The effect of smooth SFH variation is shown in the left panel.  The solid lines have exponentially decreasing SFRs with $\tau = 1.1$ Gyr, the dotted lines have more slowly falling SFRs with $\tau = 2.75$ Gyr and the dashed lines are increasing $\propto 1 - e^{-t/\tau}$ where $\tau = 1.5$ Gyr.  These tracks have solar metallicity.  (b) The effects of metallicity at fixed SFH with $\tau = 1.1$ Gyr.  $Z=0.005$ for the dashed line, 0.010 for the dotted line, 0.020 for the solid line and 0.025 for the dot-dashed line.  The solid lines are identical across the panels.  This figure demonstrates that the effects of model parameters are largely orthogonal to IMF variations.}
\label{fig:models}
\end{figure}

Within the assumption of smoothly varying SFHs much can be said about the
effects of the IMF, metallicity, age, and SFH in the color-H$\alpha$ EW plane.
Figure \ref{fig:models} demonstrates these relationships.  In both panels the
ages of the models decrease from the upper left to lower right.  The effects
of the age of a stellar population are largely orthogonal to those of IMF
variations.  In Figure \ref{fig:models}a the effects of changing the
functional form of the smoothly varying SFH with fixed metallicity and
$\Gamma$ are shown.  SFH variation is degenerate with the IMF.  However the
effect is relatively small over a wide range of SFHs.  The solid lines are
exponentially decreasing SFHs with $\tau=1.1$ Gyr where the bulk of the star
formation occurs early in the galaxy's life.  The dashed lines have SFHs that
are increasing with time where most star formation occurs at late ages.  The
effect of variations in the form of smooth SFHs is larger at later ages and
higher values of $\Gamma$ but does not dominate the effects of IMF
variations.  With all other parameters fixed the range of smooth SFHs cause systematic uncertainties at the level of $\pm 0.1$ in $\Gamma$.

In Figure \ref{fig:models}b the effects of metallicity variations with fixed
SFH and $\Gamma$ are shown.  The metallicity variations are also degenerate with IMF variations.  With all other parameters fixed between colors of $0.1<(g-r)_{0.1}<0.4$ the systematic uncertainty due to metallicity is less than 0.05 in $\Gamma$.  This  uncertainty increases to 0.35 at $(g-r)_{0.1}=-0.2$ and 0.7.

As aforementioned, the extinction correction is another potential problem.
The arrows in Figure \ref{fig:models} show the length and direction of the
extinction correction for typical galaxies in our sample.  It is assumed that
$f=2$, but $f=1$ and $f=4$ are also plotted to show the potential effect of
variations in $f$.  The reddening vectors for $f=2$ and 4 are fortuitously
orthogonal to the IMF variations.  Only when the continuum and emission
extinctions are equal, when $f=1$, do the extinction correction and variations
in $f$ become a larger concern than metallicity and SFH.  However such low $f$ ratios are not observed in galaxies (\citet{Calzetti94}  and section \ref{sec:fratio}).

\begin{figure}
\plotone{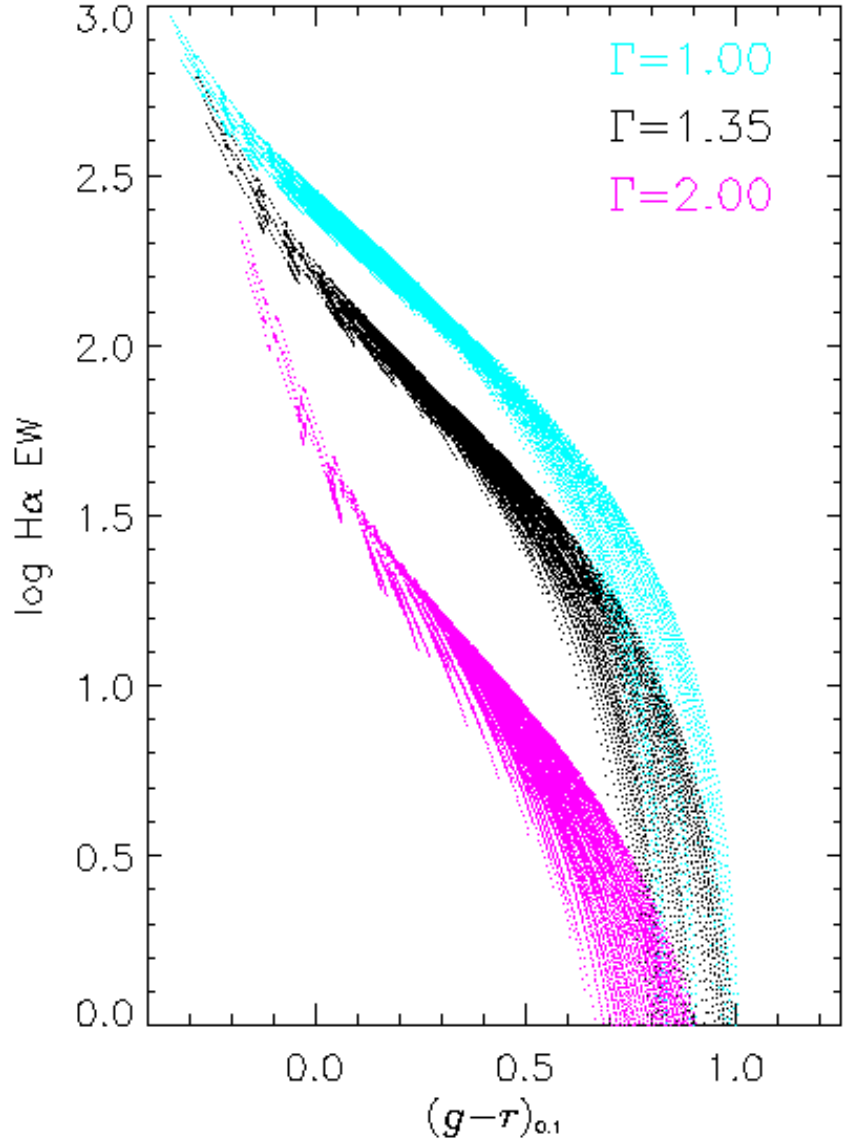}
\caption{All model points for $\Gamma=1.00$ (cyan), $\Gamma=1.35$ (black) and $\Gamma=2.00$ (magenta).  Each IMF has 18,480 calculated model values.  $Z$ ranges from 0.005 to 0.025, ages range from 100 Myr to 13 Gyr and SFHs cover the range described in the text. }
\label{fig:all135}
\end{figure}

Figure \ref{fig:all135} shows all 18,480 model points with $\Gamma=1.35$.
Models are interpolated in SFH history for fuller coverage of the
color-H$\alpha$ EW plane.  For each value of $\Gamma$ the models cover a
stripe rather than a single line.  It can be seen in the lower right of figure
\ref{fig:all135} that the model become degenerate in $\Gamma$ for old, red
galaxies with weak current star formation.

\section{Statistical Techniques}

The data and models are compared using a ``pseudo-$\chi^2$" minimization.
For various reasons (detailed below) the classical $\chi^2$ estimator
assumptions are violated so we can not use traditional tables for  
error estimates but we can still use the $\chi^2$ as a statistical estimator as long as
the confidence regions are calibrated by Monte Carlo (MC) techniques as we  
will do.  We proceed as follows:  for each galaxy $i$ we have a measured $(g-r)_{0.1}$ color, $c_i$, and an H$\alpha$ EW, $w_i$, and measurement errors $\sigma_{c_i}$ and $ 
\sigma_{w_i}$, given by equations \ref{eq:sigmac} and \ref{eq:sigmaw}.  We also have  
model values $c(\Gamma, Z,\tau,\psi)$ and $w(\Gamma, Z,\tau,\psi)$ for a  
range of IMF slopes $\Gamma$, metallicities $Z$, ages $t$ and SFHs $\psi$.  We  
can then construct a $\chi^2$ value as
\begin{equation}
\chi_i^2(\Gamma, Z,t,\psi) = \left ( \frac{c_i-c(\Gamma, Z,t,\psi)} 
{\sigma_{c_i}}\right )^2
+ \left ( \frac{w_i-w(\Gamma, Z,t,\psi)}{\sigma_{w_i}}\right )^2
\label{eq:chisquare}
\end{equation}
which is calculated by brute force.  The goal of this paper however is to investigate the IMF with relatively
simple measurements of the H$\alpha$ EW and a broadband color.  While  
making crude measurements of the mean stellar metallicities of individual  
galaxies is possible, disentangling age and SFH effects on an individual basis is a
daunting task.  Assuming that it is possible to do, it does not scale  
up well to the high redshift universe where observations will be of lower  
quality.

It does not make sense to minimize $\chi^2$ over all galaxies for a  
particular set of $(\Gamma,Z,t,\psi)$ because we have no a priori reason to  
think that all of the galaxies should have the same metallicity or SFH.  In fact we
expect that they would not.  The solution is to marginalize $\chi^2$  
over metallicity, age and SFH for each galaxy such that
\begin{equation}
\chi_i^2(\Gamma) = \min  \left [ \chi_i^2(\Gamma, Z,t,\psi) \right ]_ 
{Z,t,\psi}
\label{eq:chisquare2}
\end{equation}

This is somewhat unorthodox, because for some galaxies the data points are
over-fitted, i.e. there will be a stripe in ($c,w$) space corresponding to a
given $\Gamma$ and we can get $\chi^2_i(\Gamma)$ values very close to zero
(but not exactly because of the discrete nature of the model grid) for
galaxies within the stripe. We note we also have partial degeneracy between
parameters such as age and metallicity --- they both shift the tracks in
similar directions largely orthogonal to $\Gamma$ (though not completely which
is why we have a stripe in parameter space not a line). This makes it
difficult to calculate the traditional ``number of degrees of freedom."
Despite these limitations it is clear that galaxies inconsistent with a
particular $\Gamma$ will {\em still\/} have large values of
$\chi^2_i(\Gamma)$-- for example a very blue galaxy with a low equivalent
width in Figure \ref{fig:all135}. The complication is that the stripes for
similar $\Gamma$ values overlap, and for red galaxies with low equivalent
widths the stripes for vastly different $\Gamma$ values overlap.  As such, the
IMF for an individual galaxy is only broadly constrained.  Measuring a precise
best IMF for an individual galaxy boils down to random chance and the discrete
nature of the models.  However, by summing $\chi_i^2$ over many galaxies the
IMF is narrowly constrained for the sample being summed over as long as we are
careful in our confidence region analysis. 

Because of this over-fitting and partial degeneracy we can not apply the textbook
notions of the  $\chi^2$ distribution, calculate degrees of freedom
and choose $\Delta\chi^2$ contours for different confidence
regions. Further to this $c_i$ and $w_i$ are not truly independent  
variables.  Both the colors and EWs are subject to the same extinction and reddening
corrections which tie the errors together.  For galaxies with $z
\lesssim 0.04$ the H$\alpha$ line is in the observed $r$ band,  
although this only affects a relatively small number of galaxies in the sample  
almost all of which have $M_{r,0.1} > -20$.  Also the direct statistical interpretation of $\chi^2$ is predicated on the assumption of normal errors.  Equations \ref{eq:sigmac}, \ref 
{eq:sigmaav} and \ref{eq:sigmaw} reveal that our errors are complicated mixtures of  
individual measurements which are most likely Poisson distributed.  Thus $\sigma_c$ and
$\sigma_w$ are unlikely to be normally distributed.  Bursty SFHs can
potentially create outliers which are statistically significant due to the
fact that neither $\sigma_c$ or $\sigma_w$ include a term for this difficult
to quantify effect.  The problem is even worse if the errors are non-symmetric
which could potentially arise from the aforementioned bursty SFHs.  In the
case of non-symmetric errors the best value of $\Gamma$ could erroneously be
pulled away from the true value.

Given all this we abandon the direct statistical interpretation of $\chi^2$ and regard it as an estimator of the goodness of fit whose confidence regions have to be calibrated empirically. We do this via MC simulations (as recommended by \citet{Press}) where we simulate data points for a given $\Gamma$ with the correct error distributions and propagate everything through the analysis in the same way as for  
the actual data.  For each of our MC simulations we add Poisson errors to the $ugriz$  
fiber magnitudes, H$\alpha$ and H$\beta$ fluxes and the observed H$\alpha$  
EW.  We assume that these observed quantities have Poisson dominated errors-- as
members of the MGS they are high S/N measurements.  The entire analysis
described above is repeated, including a new extinction measurement  
and a recalculation of the $K$-corrections.  For each value of $\Gamma$ we run 100 MC simulations to estimate the 95\% confidence interval. Setting up the MC architecture in this  
way has the further advantage that we can use the same machinery to test the effect of systematic
errors such as the violation of our smooth SFH assumptions, as we will do later.
The main downside of course is that this approach is computationally intense. Run times
for the 100 MC simulations are typically 18 days on a 2 Ghz desktop PC for the samples  
considered here.

In practice it turns out that $\chi^2(\Gamma)$ is still a smooth well-behaved function with,
not surprisingly, a quadratic minimum which has the advantage that we  
can then interpolate it to increase the resolution in the
best-fitting $\Gamma$ without incurring the additional computational  
expense. This arises of course from the fact that our estimator is similar to a  
traditional $\chi^2$ and is a good reason to stick with this similarity over some more exotic goodness  
of fit measure.  An estimate of the systematic errors is discussed later.

Regardless of how poorly a sample is modeled by a universal IMF the  
above method will still find a best fitting $\Gamma$ and corresponding  
confidence region.  We still expect $\chi^2$ to be small for a model that is a good fit and large for one that is not.  One nuance in comparing $\chi^2$ between different sub-samples, as we will do, is  
that the samples are often of considerably different sizes. Because of this we  
choose instead to use the {\em mean} $\chi^2$, $\overline{\chi^2}$ instead, as a sample
metric. This has the advantage that absolute $\overline{\chi^2}$ values and confidence regions are more similar between the sub-samples, though we note that the confidence regions on
$\overline{\chi^2}$ are still determined directly from our MC simulations.

\section{Monte Carlo Results}

Figure \ref{fig:all-like} shows the results of our analysis for the full
sample of galaxies using just the observed data set.  The ``X'' marks the best
fitting IMF, where $\Gamma=1.4411$ and $\chi^2 = 60549.7$ with
$\overline{\chi^2} = 0.46$.  At $\Gamma=1.00$ $\overline{\chi^2} = 3.61$ while
steeper IMFs are more heavily rejected with $\overline{\chi^2} = 9.00$ at
$\Gamma=2.00$.

For comparison several ``classic'' IMFs are also plotted in figure
\ref{fig:all-like} at their approximate equivalent values of $\Gamma$.  With
$\overline{\chi^2} = 7.73$ the \citet{MS79} solar neighborhood IMF is a
particularly bad fit.  Two more recent solar neighborhood IMFs,
\citet{Scalo86} and \citet{Kroupa93}, yield $\overline{\chi^2} = 2.38$ and
1.98 respectively.  These results reinforce the conclusions of K83, KTC94 and
\citet{BG03} that the solar neighborhood is not representative of galaxies
on the whole as far as the IMF is concerned.

On the other hand the \citet{Scalo98} IMF, established from a review of star
cluster IMF studies in the literature, is a better fit than the best $\Gamma$
value in our parameterization with $\overline{\chi^2} = 0.43$.  This result
highlights the degeneracy of the IMF models themselves in the color-H$\alpha$
EW plane-- two considerably different IMFs (one with one break and the other
with two) fit nearly equally as well.

The results of the MC simulation show that the 95\% confidence region is
$1.4432<\Gamma<1.4443$ for the data set as a whole.  The MC simulation shows
that additional data will not improve the overall results as the random errors
are already small.  Clearly and not surprisingly systematic errors, which are
discussed later, dominate.

The overall result of $\Gamma=1.4437 \pm 0.0005$ is steeper than the original
Salpeter value of $\Gamma=1.35$.  It is also steeper than the \citet{BG03}
value of $\Gamma = 1.15\pm0.2$ derived from galaxy luminosity densities in the
UV to NIR.  It is however well within their 95\% confidence limit of
$\Gamma<1.7$ as well as their measurement of $\Gamma=1.2\pm0.3$ based on the
H$\alpha$ luminosity density.  The difference between their two results
suggests that the H$\alpha$ and mid-UV to optical fluxes may have different
sensitivities to massive stars.  \citet{Scalo98} estimated the uncertainty in
$\Gamma$, either due to measurement uncertainties, real IMF variations or
both, in his star cluster IMF based on the spread of results in the
literature.  Our result is well within his range of uncertainty in both mass
regimes-- $\Gamma=1.7\pm0.5$ for $1-10\ {\rm M}_\odot$ and $\Gamma=1.3\pm0.5$
for $1-100\ {\rm M}_\odot$.

\subsection{Luminosity Effects}

\begin{figure}
\plotone{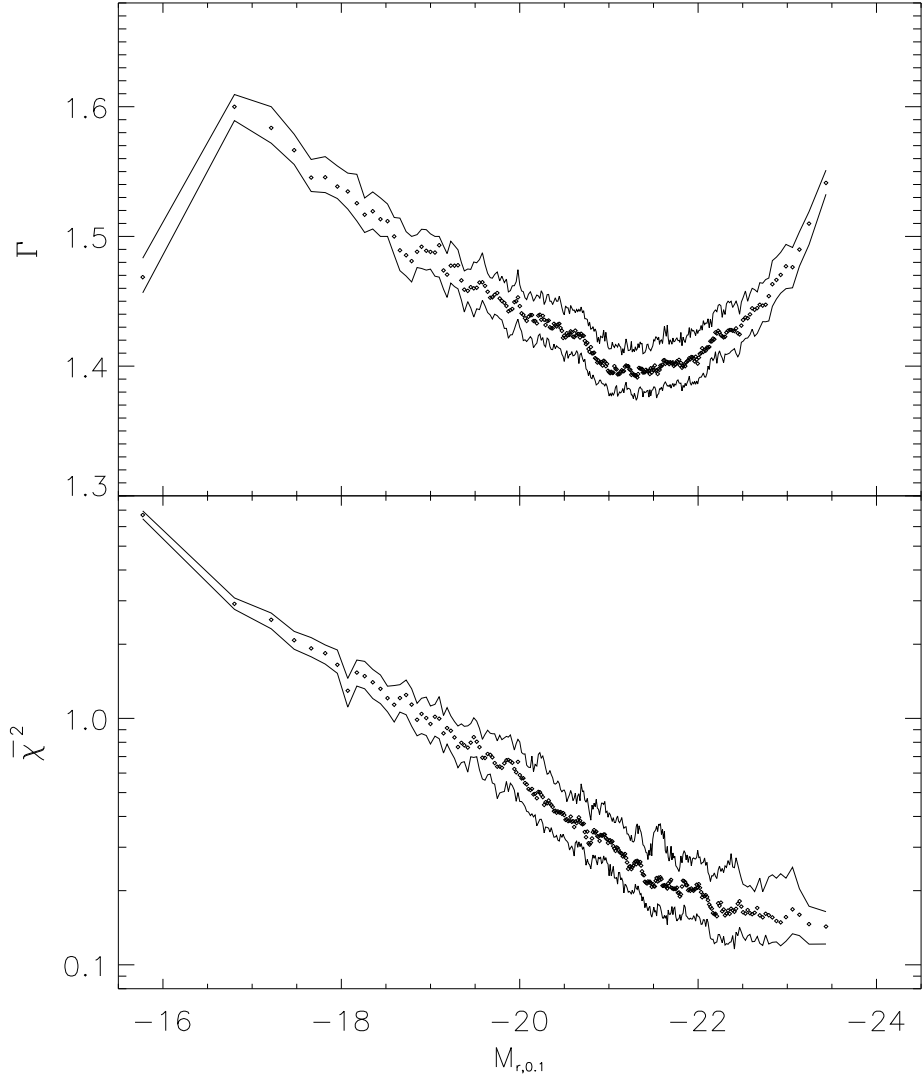}
\caption{MC simulation results for the full sample binned by ${\rm M}_{r,0.1}$.  Each diamond represents 500 galaxies plotted at the mean ${\rm M}_{r,0.1}$ value of the bin.  The points are the most dense around ${\rm M}_{r,0.1}=-21.5$ as the sample is dominated by galaxies in this luminosity regime.  The solid lines represent the upper and lower 95\% confidence region measured for each bin.  Top panel: The best fitting $\Gamma$ values as a function of $r$-band luminosity.  Lower panel:  The $\overline{\chi^2}$ values for each luminosity plotted on a log scale.}
\label{fig:allmagr}
\end{figure}

The luminosity of a galaxy could potentially have an effect on the IMF within
it.  For one the ambient radiation field is likely higher in more luminous
galaxies.  Figure \ref{fig:allmagr} shows the best fitting IMF and
$\overline{\chi^2}$ values as a function of ${\rm M}_{r,0.1}$ for all 130,602
galaxies.  The galaxies have been binned in ${\rm M}_{r,0.1}$ such that there
are 500 objects in each bin.  The bin size was chosen to maximize coverage in
${\rm M}_{r,0.1}$ yet still keep the random errors in each bin small.  The
solid lines represent the lower and upper 95\% confidence region determined
from the MC simulation.

Figure \ref{fig:allmagr} reveals a constant value of $\Gamma \sim 1.4$ for
galaxies with ${\rm M}_{r,0.1}$ between $-21$ and $-22$ with linear increases in
$\Gamma$ for both brighter and fainter galaxies.  There is also a sudden
downturn in $\Gamma$ values for galaxies fainter than ${\rm M}_{r,0.1}=-16.5$.
Given the sizes of the random errors the differences in $\Gamma$ between ${\rm
  M}_{r,0.1}=-17$ and $-22$ are substantial, from 1.59 to 1.41, and
statistically significant.  The agreement with the Salpeter slope is
the best for galaxies between $-21$ and $-22$ in ${\rm M}_{r,0.1}$.

In many ways it is not surprising that previous investigators have not found
this trend.  The Milky Way is thought to have a luminosity of ${\rm M}_V =
-20.9$ \citep{Delhaye65}; the $V$ and $r_{0.1}$ filter curves cover roughly
the same wavelengths.  At comparable luminosities our results are similar to
Salpeter.  The galaxies in the K83 sample have a median ${\rm M}_B =-20.9$
with only 16\% (18 objects) fainter than ${\rm M}_B^*=-19.7$ \citep{EEP88}.
This is a significant bias toward more luminous galaxies where our results are
in agreement with a universal IMF.  By contrast 30\% of our sample is fainter
than ${\rm M}_{r,0.1}^* = -20.44$ \citep{Blanton03b}.  We have a sample of
39,350 galaxies fainter than $\rm{L}^*$.

The lower panel of Figure \ref{fig:allmagr} shows that the relative quality of
the fits rapidly deteriorates as the luminosity of the galaxies decrease.  For
the brightest galaxies $\overline{\chi^2}$ floats around 0.15, while in the
faintest bin it is over 6.  For comparison ${\rm M}_V = -18.5$ for the Large
Magellanic Cloud and $-17.1$ for the Small Magellanic Cloud
\citep{Courteau99}.  This trend could indicate that a universal IMF is a good
fit to the most luminous galaxies, but dwarf galaxies cannot be described by a
universal IMF, even if a different universal slope is allowed.  However it
could have a more mundane explanation.  It could be that errors are over or
underestimated as a function of luminosity.  It also could arise from
deviations from our assumption of smoothly varying SFHs.

We cannot bin our data by stellar mass without assuming an IMF which is
contrary to the goals of the project.  We can repeat the analysis of Figure
\ref{fig:allmagr} using ${\rm M}_{z,0.1}$ in the place of ${\rm M}_{r,0.1}$.  
${\rm M}_{z,0.1}$, being redder, is a better proxy to stellar mass.  The
resulting plot is nearly identical to Figure \ref{fig:allmagr} which shows
that the relationship persists across several wavebands.

Figure \ref{fig:allmagr} reveals a clear, statistically significant trend in
$\Gamma$ and $\overline{\chi^2}$ with respect to luminosity.  The rest of this
section focuses whether this trend is a manifestation of true IMF variations
or if it is  the result of sample biases or poor assumptions.

\subsection{Sources of Bias}

If the IMF is truly universal and our method successfully probes the IMF any
subsample of galaxies that we could choose should yield the same $\Gamma $
value as any other in spite of any selection biases or aperture effects.  Figure \ref{fig:allmagr} clearly shows that the preceding
statement is false.  In this section we set aside the possibility of IMF
variations and search for biases in our sample.

\subsubsection{Magnitude Limited Sample}

Figure \ref{fig:allmagr} shows that the overall result of $\Gamma=1.4437$ is
really a weighted average.  The SDSS MGS is a magnitude limited
sample, one defined by flux limits, with both upper and lower limits.  Table
\ref{tab:lumbins} gives the number of objects in each luminosity bin.  There
are 41,411 galaxies with $-21.5<{\rm M}_{r,0.1}<-20.5$ but only 28 for which
$-14.5<{\rm M}_{r,0.1}<-13.5$.  As such the overall result is heavily biased
by more luminous galaxies.

Malmquist bias will affect any magnitude limited sample.  Because brighter
objects can be seen at greater distances a magnitude limited sample contains
bright objects from a greater volume of space than fainter objects.  The
result is that the ratio objects by luminosity in a magnitude limited
sample differs from the true ratio in nature; brighter objects are
over-represented.

\begin{figure}
\plotone{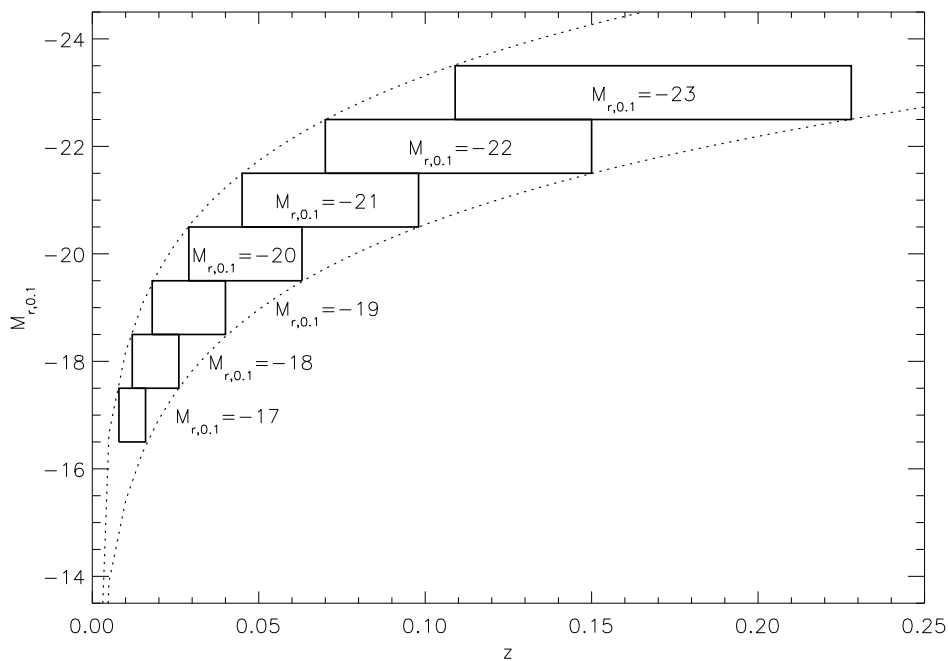}
\caption{Definition of volume limited magnitude bins.  The dotted lines show
  the luminosities corresponding to the flux limits of the SDSS MGS ($15.0 < r
  < 17.77$) as a function of redshift.  The solid boxes are the volume limited
  magnitude bins used in this paper.  Within each box no galaxies within the
  magnitude range of the box are affected by the flux limits of the sample.}
\label{fig:VL-def}
\end{figure}

To eliminate Malmquist bias volume limited bins where subsamples are complete for a range of luminosities are constructed.  Figure
\ref{fig:VL-def} details the construction of these bins.  Given both the
upper and lower flux limits of the MGS ($15.0 < r < 17.77$) only a factor of
13 in luminosity falls in the sample at any given redshift.  The redshift
limits of each volume limited bin are defined such that no galaxies within the
magnitude limits of the bin are affected by the flux limits of the MGS.
Within each box in Figure \ref{fig:VL-def} the true ratio of galaxy
luminosities is preserved and is thus free of Malmquist bias.

\begin{figure}
\plotone{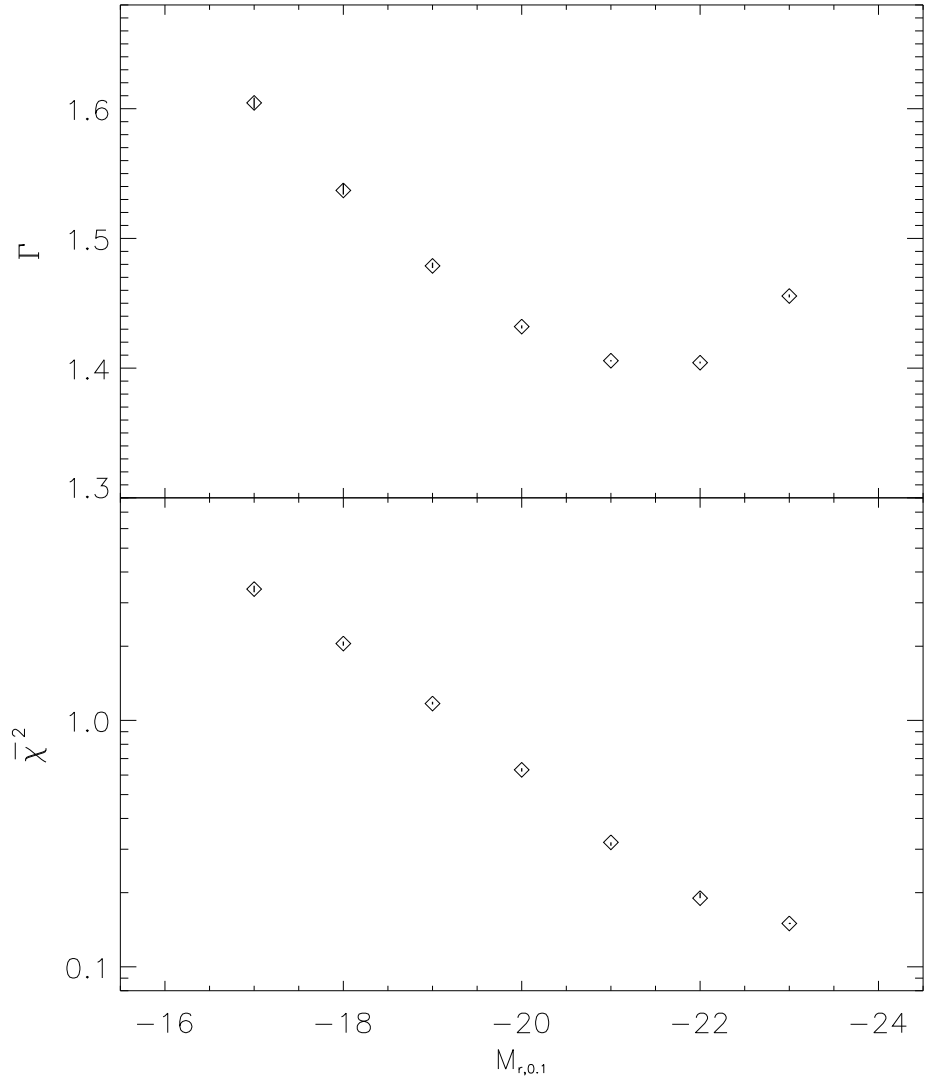}
\caption{MC simulation results as a function of ${\rm M}_{r,0.1}$ where the magnitude bins are volume limited.  The error bars represent the 95\% confidence interval from the simulation and are in most cases smaller than the plotting symbol.  Top panel: The best fitting $\Gamma$ values as a function of $r$-band luminosity.  Lower panel:  The $\overline{\chi^2}$ values for each luminosity plotted on a log scale.}
\label{fig:VL-plot}
\end{figure}

Figure \ref{fig:VL-plot} shows the results for volume limited magnitude bins.
Most error bars are smaller than the plotting symbols due to the larger number
of galaxies, 329 to 29,701 as given in Table \ref{tab:lumbins}, in each bin.
Figures \ref{fig:allmagr} and \ref{fig:VL-plot} show the exact same trends.
The largest difference in $\Gamma$ between the whole and volume limited
samples is 0.0116 in the ${\rm M}_{r,0.1}=-17$ bin.  The other notable
difference is that the fainter galaxies have larger $\overline{\chi^2}$ values
in the volume limited case.  However Malmquist bias across bins is not
responsible for the luminosity trends in Figure \ref{fig:allmagr}.

\subsubsection{Redshift}

Another effect of magnitude limited samples is that the faintest galaxies are
much closer than the most luminous ones.  The mean redshifts of the volume
limited magnitude bins range from $z=0.013$ for ${\rm M}_{r,0.1}=-17$ to
$z=0.168$ for ${\rm M}_{r,0.1}=-23$.  This corresponds to a difference in age
of around 1.8 Gyr.  As aforementioned, model tracks reveal that age is largely
orthogonal to the IMF in our parameter space, but there could be other effects
tied to age and distance.  In addition the IMF could evolve with time.

The large number of galaxies in our sample affords us the luxury of
investigating the effects of luminosity and redshift simultaneously to obtain
a better understanding of what role, if any, the redshift plays in our
analysis.  Figure \ref{fig:redshift} shows our fitted parameters for all
130,602 galaxies in bins that are 0.25 magnitudes wide in luminosity and 0.005
wide in redshift.  The upper left panel shows the best fitting $\Gamma$ for
each two dimensional bin.  On the upper right the width of the 95\% confidence
region in $\Gamma$ for each bin is shown.  At bottom left is the $\log
\overline{\chi^2}$ and at bottom right is the log of the number of galaxies in
each bin.  The white contour demarcates the region in which each bin contains
at least 50 galaxies.  The number 50 is arbitrary but it shows the region
where Poisson errors are expected to be small.  The black areas are regions
where there are no galaxies with the given parameters.

Using the plot of $\Gamma$ at the upper left we can look for potential
redshift biases.  This is complicated by the fact that at a fixed luminosity
there is a limit to the range of redshifts in the sample due to the flux
limits of the sample described earlier.  Looking at vertical slices through
the plot at any fixed luminosity there is a trend towards larger values of
$\Gamma$ with increasing redshift.  However, for horizontal slices of fixed
redshift the same relationship between $\Gamma$ and luminosity that is present
for the whole sample is seen modulo a normalization factor.

The right half of Figure \ref{fig:redshift} shows a strong relationship between
the number of galaxies per bin and the width of the 95\% confidence region in
$\Gamma$.  This simply reflects the fact that larger samples are less affected
by Poisson errors.

The lower left panel of Figure \ref{fig:redshift} provides an excellent
example of why our metric of fit quality, $\overline{\chi^2}$, is so
important.  Bins with similar numbers of galaxies and $\delta \Gamma$ values
can have vastly different values of $\overline{\chi^2}$.  It is worth
reminding that the contours in $\overline{\chi^2}$ are logarithmic.  At
fixed luminosity the galaxies are better fit by a universal IMF at higher
redshift.  Similar to the sample as a whole the quality of fit improves with
luminosity.

While there does appear to be some weak trending of $\Gamma$ and
$\overline{\chi^2}$ with redshift, redshift effects are not driving the
relationship seen between IMF and luminosity as it persists at fixed redshifts.
 
\subsubsection{Aperture Effects}

One explanation for the trend in $\Gamma$ with redshift is aperture effects.  Again, if the IMF is truly universal aperture effects should not exist.  
The SDSS spectra have a fixed aperture of 3'' for all galaxies.  Depending on
the angular extent and distance to a galaxy a different fraction of the total
light of the galaxy will fall into the aperture.  The problem is mitigated by
the fact that the most distant galaxies are the most luminous and more likely
to have a larger physical size.  As the physical area contained in the aperture
increases with distance, so too does the size of the galaxies being observed.
However, the two effects do not exactly balance out.  Table \ref{tab:lumbins}
shows that the mean aperture fraction for the ${\rm M}_{r,0.1} = -17$ bin is
0.20 and increases to 0.27 at ${\rm M}_{r,0.1} = -23$.  On average 35\% more
of the most luminous galaxies fall within the aperture compared to the faintest.

Figure \ref{fig:aperture} shows the behavior of our fitted parameters for two
dimensional bins of luminosity and aperture fraction in the same manner as
Figure \ref{fig:redshift} did for luminosity and redshift.  For fixed
luminosities increasing aperture fraction leads to decreasing values of
$\Gamma$.  However at fixed aperture fraction the qualitative IMF-luminosity
trend remains.  The $\overline{\chi^2}$ values are a strong function of luminosity,
but $\overline{\chi^2}$ does increase with aperture fraction at each fixed luminosity.

The trend with aperture fraction is the exact opposite of what would be
expected in the presence of a systematic effect operating given the redshift result in Figure \ref{fig:redshift}.  The nearest galaxies should have the smallest aperture fractions in a particular
luminosity bin.  The nearest galaxies in Figure \ref{fig:redshift} have the
smallest values of $\Gamma$ while the smallest aperture fractions in Figure
\ref{fig:aperture} have the largest values of $\Gamma$.

Figure \ref{fig:aperture} suggests that the measured IMF is more dependent on
the aperture fraction than the redshift.  There are several possible physical
explanations for IMF trends with the aperture fraction, all of which are
related to radial gradients in disk galaxies.  \citet{Padoan97} make the
theory based claim that the IMF should be a function of the original local
temperature of the star-forming molecular clouds.  Metallicity gradients are
also known to exist in disk galaxies, including the Milky Way \citep{Mayor76}.
\citet{Rolleston00} measure a linear, radial light metal (C, O, Mg \& Si)
abundance gradient of $-0.07 \pm 0.01 \rm{\ dex\ kpc}^{-1}$ in the disk of the
Milky Way.  Given the increased efficiency of cooling with metal lines we
would expect the most low mass stars where metallicity is the highest-- on
average towards the center of galaxies.  The trend in $\Gamma$ in Figure
\ref{fig:aperture} is qualitatively consistent with this idea.

If there are radial IMF gradients in galaxies one would expect the fits to
decrease in quality with increasing aperture fraction.  A blend of IMFs will
not be fit as well as a universal one given our technique.  This idea is
consistent with the results in Figure \ref{fig:aperture}.  However, in
well-resolved stellar populations there is no evidence for a relationship
between the IMF and metallicity, except perhaps at masses lower than those
probed by our method \citep{Kroupa02}.  If metallicity plays a role in
determining the IMF the effects are only being revealed as a global trend in
our large sample of integrated stellar populations.  For individual clusters
metallicity must play a secondary role to stochastic effects.

\subsubsection{Extinction Correction}

The extinction correction is another potential source of bias.  It is possible
that there is a second order correction that our fairly simple extinction
correction fails to take into account.  This could potentially lead to an
erroneous IMF trend with extinction correction.  This affects the luminosity
results because more luminous galaxies tend to be dustier, as evidenced in
Table \ref{tab:lumbins}.  The problem is further complicated by the fact that
dust is thought to play an integral part in star formation so it is not
unreasonable that an observed IMF trend with extinction may be real.

Figure \ref{fig:extinction}, similar to Figures \ref{fig:redshift} and
\ref{fig:aperture},  shows the results of our analysis for two dimensional
bins of luminosity and the extinction correction that was measured and
applied.  Vertical slices through the upper left panel of Figure
\ref{fig:extinction} show that $\Gamma$ does depend on ${\rm A}_{{\rm V},l}$,
trending towards lower $\Gamma$ values with increased extinction over the
region where the Poisson error in $\Gamma$ is reasonable.  Yet again,
horizontal cuts of fixed extinction show the IMF-luminosity relationship.

The decreasing $\Gamma$ values with increasing extinction is
counter-intuitive.  Dustier regions tend to be more metal rich.  If
metal cooling plays a significant role in the IMF the dustiest regions
should have the steepest IMFs.

At fixed luminosity $\overline{\chi^2}$ increases with extinction.  As
aforementioned our calculated errors in color and EW (equations
\ref{eq:sigmac} and \ref{eq:sigmaw}) have a functional dependence on the
observed emission line extinction.  In both cases it is the term proportional
to $\sigma_f {{\rm A}_{{\rm V},l}}$ which is on average the major contributor
to the calculated error.  Because luminous galaxies tend to be more heavily extincted they
will also be more likely to have larger errors.  This is potentially
problematic for our observed IMF trend with luminosity.  If we are unknowingly
underestimating the errors for faint galaxies with low extinction the source
of the poor fit qualities of these galaxies could be systematic instead of
astrophysical.  However the lower left panel of Figure \ref{fig:extinction}
shows that the most extincted galaxies have the poorest fits where such a bias
would suggest that they should fit the best due to the large accommodating errors.

\subsubsection{Multiple Parameter Biases}

It is also possible that biases in our $\Gamma$ measurements could depend on
two parameters simultaneously.  Figure \ref{fig:avap} shows the measured
galaxy of $\Gamma$ as a function of both aperture fraction and measured
emission line extinction for six volume limited luminosity bins.

When holding all other parameters fixed, increasing the aperture fraction
leads to lower values of $\Gamma$ in all statistically significant areas of
Figure \ref{fig:avap}.  This is the same relationship found in the earlier
section on aperture fraction.  Decreasing values of $\Gamma$ are also seen for
increasing extinction when all other parameters are constant.  A notable
exception to this is that galaxies with large extinction and small aperture
fractions favor higher $\Gamma$ values.

Most importantly when looking at a particular combination of aperture fraction
and extinction the IMF becomes shallower with increasing luminosity until the
highest luminosities where it becomes steeper again.  Even in the narrowest
slices of the data set the same IMF-luminosity trend is seen, albeit with
slightly different absolute values of $\Gamma$.

\subsubsection{Star Formation Strength}

As discussed previously we have allowed two classes of star forming galaxies
into our sample.  111,806 galaxies (86\%) fall in the star forming class and
the other 18,796 (14\%) belong to the low S/N star forming class where the
\ion{O}{3} or \ion{N}{2} lines are weak, but the H$\alpha$ and
H$\beta$ lines still have S/N $>5$.  By comparing the results from
these two subsamples we can investigate a possible bias of the results
with respect to the level of star formation.

\begin{figure}
\epsscale{0.75}
\plotone{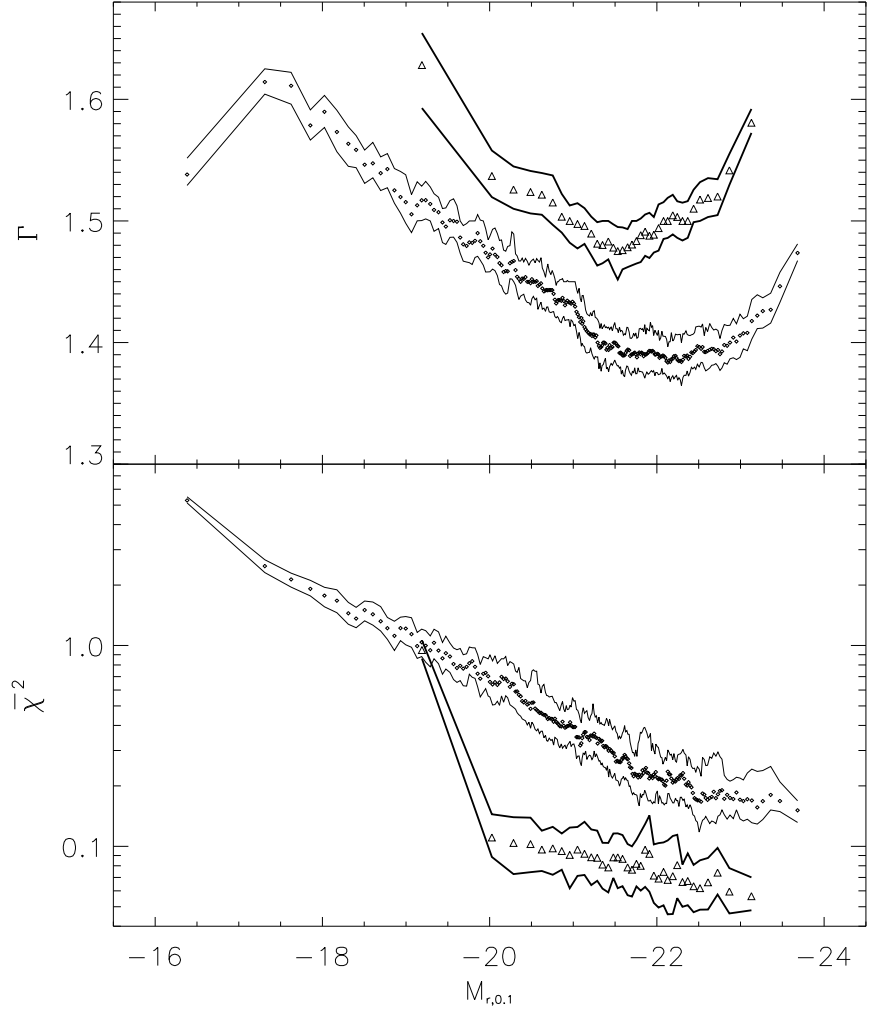}
\caption{MC simulation results binned by ${\rm M}_{r,0.1}$ for the 111,806
  galaxies in the star forming class (diamonds) and the 18,796 galaxies in the
  low S/N star forming class (triangles) of \citet{Brinch04}.  Each symbol represents
  a bin of 500 galaxies.  The thin lines represent the upper and lower 95\%
  confidence region measured for each star forming bin, while the thick lines
  do so for the low S/N bins.  Top panel: The best fitting $\Gamma$ values as
  a function of $r$-band luminosity.  Lower panel:  The $\overline{\chi^2}$
  values for each luminosity plotted on a log scale.  Because the sample is
  dominated by the star forming class the results for these galaxies is very
  similar to the result for the whole sample in Figure \ref{fig:allmagr}.}
\label{fig:sfclass}
\end{figure}

Figure \ref{fig:sfclass} shows the results for both classes as a function of
luminosity.  As it comprises 86\% of the total sample it is not surprising
that the results for the star forming class are similar to those of the sample
as a whole in Figure \ref{fig:allmagr}.

The low S/N class exhibits a similar qualitative behavior to the set as a
whole with a few notable differences.  The $\Gamma$ values are offset by at
least 0.08 towards larger $\Gamma$.  The measured IMF turns toward steeper
values at lower luminosities than for the sample as a whole.  The
$\overline{\chi^2}$ are several times lower as well.

While the galaxies in low S/N class meet the same requirement of S/N $>$ 5 in
the Balmer lines as the star forming class they are biased towards noisier
H$\alpha$ line measurements.  This corresponds to lines that are either weak
(low SFR) or weak compared to the continuum (low present SFR compared to the
past) both of which lead to low EWs.  Another issue at play is that
the relationship between the H$\alpha$ line flux and the SFR is
dependent on the IMF.  At a fixed metallicity and SFR increasing
$\Gamma$ by 0.05 reduces the H$\alpha$ flux by 20\%.  In fact the
H$\alpha$ flux of a galaxy with $\Gamma=1.00$ will be 33 times larger
than a galaxy with the same SFR and $\Gamma=2.00$.  In the presence of real IMF variations at any fixed luminosity the low S/N class will be biased towards galaxies with steeper IMFs.  Both low SFRs and steep IMFs potentially lead to low S/N H$\alpha$ flux.  However it is difficult to determine the level of influence of each effect.

As shown in figure \ref{fig:grha_data} low H$\alpha$ EWs lead to larger values of $\Gamma$ for any fixed color.  As the low S/N class tends toward noisier H$\alpha$ fluxes and therefore EWs it is easy to see from equations \ref{eq:sigmac},
\ref{eq:sigmaav} and \ref{eq:sigmaw} that the errors for this class will tend
to be larger.  This in turn leads to lower $\chi^2$ values.

As the qualitative IMF-luminosity trend occurs in both star forming classes
the strength of star formation is unlikely to be a significant bias on our results.

\subsubsection{The $f$ Ratio}\label{sec:fratio}

As mentioned in the data corrections section, the $f$ ratio is ratio of the
extinction experienced by the nebular emission lines to that experienced by the
stellar continuum.  The assumption of a value for $f$ could potentially bias
our results.  An alternative way of looking at the same problem is that our
data in the color-H$\alpha$ EW plane can be used to constrain the $f$ ratio by
assuming a universal Salpeter IMF.

Figure \ref{fig:f-all} gives the results of this analysis for the data set as
a whole.  A value of $f=2.0$ is found with $\overline{\chi^2}=1.068$.  This is
in good agreement with to the \citet{Calzetti94} value of $f=2.0+0.6/-0.4$.
The quality of the fit in the best case is worse than in Figure
\ref{fig:all-like}.  Part of the reason for this is that the errors used were
slightly smaller as the $\sigma_f$ terms in equations \ref{eq:sigmac} and
\ref{eq:sigmaw} are set equal to 0.  The quality of the fit drops sharply
below $f=2$ and more gradually for $f>2$.  Values of $f$ near 1 are heavily
rejected.  However in this particular plot the results are dominated by
luminous galaxies.

\begin{figure}
\plotone{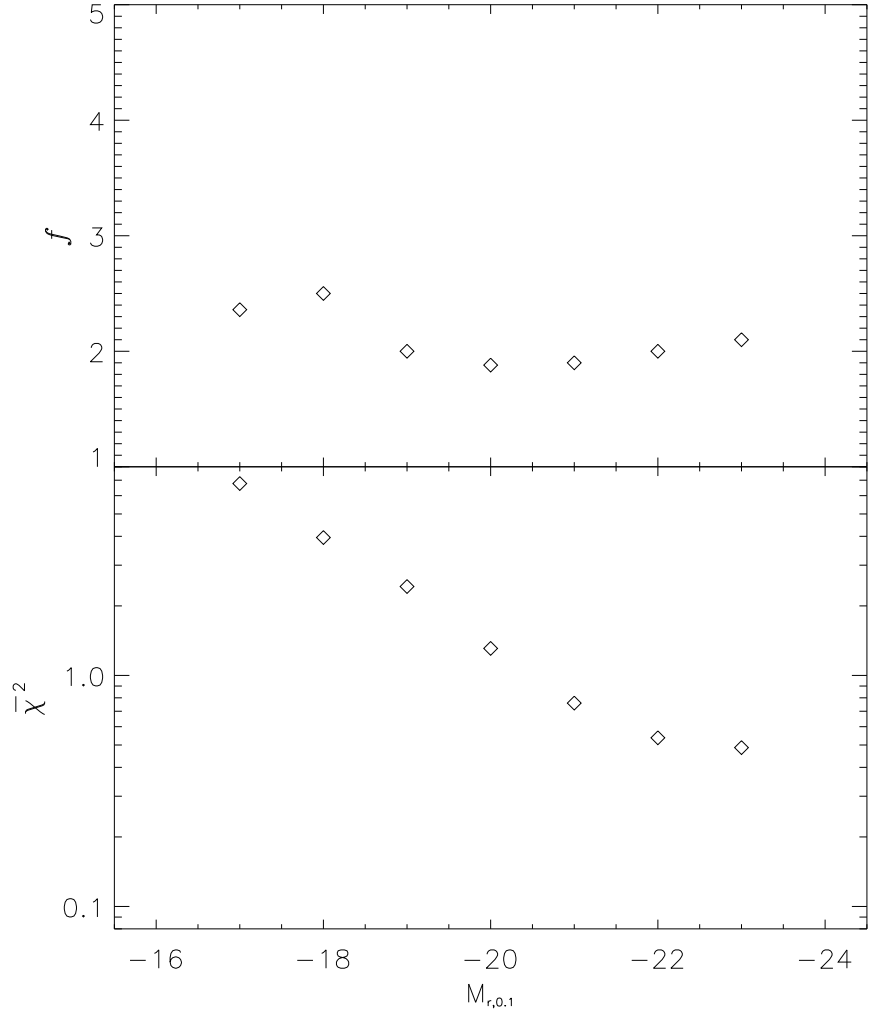}
\caption{Upper panel: The best fitting $f$ ratio, assuming a universal Salpeter IMF, as a function of luminosity.  The luminosity bins are volume limited and the same as in Figure \ref{fig:VL-plot}.  Lower panel: The corresponding modified $\overline{\chi^2}$ values as a function of luminosity.}
\label{fig:f-VL}
\end{figure}

The values of $f$ as a function of luminosity are shown in Figure
\ref{fig:f-VL}.  For galaxies ${\rm M}_{r,0.1}=-19$ and brighter the best
value of $f$ is consistent with $f=2$.  The faintest two bins the prefer an
$f$ ratio closer to 2.5.  However the lower panel shows that this new $f$ value
does not translate to improved fit quality.  In fact the faintest galaxies
have in general smaller measured extinctions and are therefore less
susceptible to changes in $f$.  The same qualitative trend of worsening fits
with decreasing luminosity seen when allowing $\Gamma$ to float is seen with a
varying $f$ value.

Together these two $f$ ratio plots provide a number of insights.  For one it
shows that our choice of the $f$ ratio is very sensible and provides an
independent confirmation of other $f$ ratio measurements.  The fact that our
best fitting $\Gamma$ values are at least 0.05 above the Salpeter value cannot
be reconciled by changing the geometry of the dust screen.  It provides
further evidence that the relationship between $\Gamma$ and luminosity is not
a function of extinction or a byproduct of our extinction correction.

\subsubsection{Summary}

In this section we have investigated several possible sources of bias to
account for our observed trend between the IMF and luminosity.  Relationships
between the IMF and redshift, aperture fraction, extinction and star
formation strength have been uncovered.  Two parameter biases were also
found.

In all cases in narrow slices through the data where potential biases are held
fixed the qualitative IMF-luminosity relationship appears.  The parameters
primarily act to offset the value of $\Gamma$ at a particular luminosity.  The
ratio of continuum to emission line extinction, $f$, was found to be a
sensible choice and the results are not sensitive to small changes in this
value. 

There are two possible interpretations to the relationships between the IMF
and potential biases.  One is that they are systematic effects due to some
problem with our measurement of $\Gamma$.  The second is that they are real
physical effects.  It is not clear from the data which of these statements is
more correct.

\subsection{Star Formation History}

\begin{figure}
\plotone{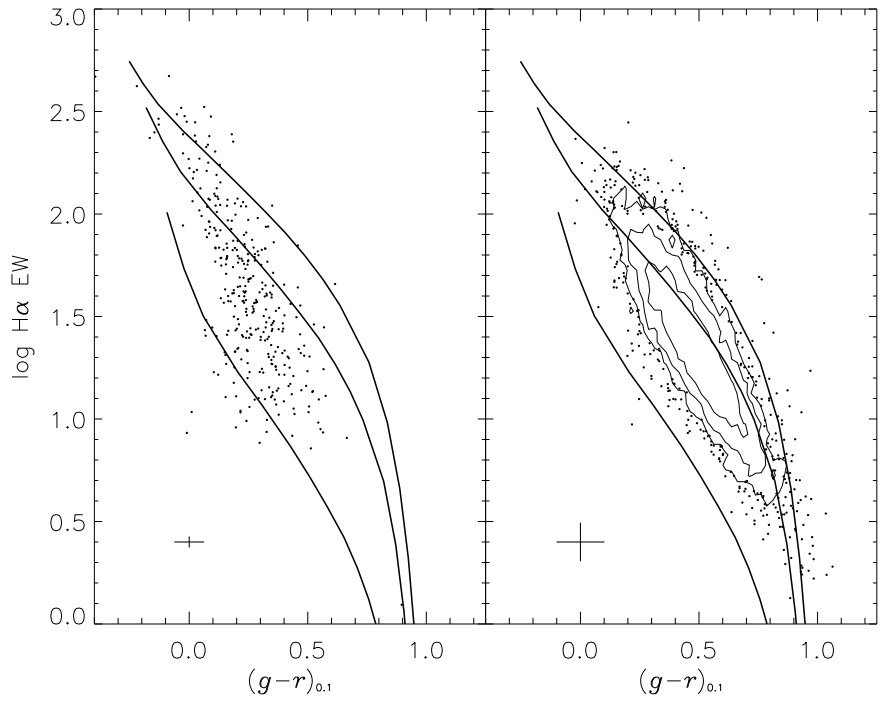}
\caption{Distribution in $(g-r)_{0.1}$ -- $\log(\rm{H}\alpha \rm{\ EW})$ space of the 329 galaxies in the volume limited ${\rm M}_{r,0.1}=-17$ bin (left) and the 8,049 galaxies in the ${\rm M}_{r,0.1}=-23$ bin (right).  The contour levels and other descriptions are identical to Figure \ref{fig:grha_data}.}
\label{fig:hagr1723}
\end{figure}

In the previous section several possible sources of bias were investigated, but
none were able to account for our observed trend in $\Gamma$ with luminosity or
the inability of a universal IMF to fit low luminosity galaxies.  Figure
\ref{fig:hagr1723} shows the distribution in color-H$\alpha$ EW space for the
least and most luminous bins,  ${\rm M}_{r,0.1}=-17$ and $-23$.  From this
figure it is apparent that the most luminous galaxies lie roughly parallel to
the IMF tracks while the faintest galaxies are more perpendicular to the
tracks.  In the low luminosity bin there are galaxies which are simultaneously
blue and have low EWs.  These galaxies are not consistent with a universal IMF
with $\Gamma=1.35$ and as mentioned before are not consistent with a universal
IMF with a different slope.  In addition the faintest galaxies have the lowest extinctions they are the least sensitive to dust and $f$ ratio issues.  Before concluding that this is evidence for IMF
variations we must first consider whether our model assumption of smooth SFHs is justified.

\subsubsection{Effects of Star Formation Bursts}

The SFH of individual galaxies is the most problematic aspect of the K83
analysis.  A sudden burst on top of a smoothly varying background will
immediately increase the H$\alpha$ EW.  This is due to the formation of O and
B stars which indirectly increase the H$\alpha$ flux through processing of
their ionizing photons.  The new presence of O and B stars also makes the
color of the galaxy bluer.  Both of these effects are proportional to the size
of the burst.

After the burst is over the H$\alpha$ EW is smaller and the colors are redder
than they would be if the burst had not occurred.  The H$\alpha$ EW drops
because there is no longer an excess of O and B stars and their ionizing
photons which reduces the H$\alpha$ flux to pre-burst levels.  However there
is now an excess of red giants due to the less massive stars from the burst
leaving the main sequence.  This increases the continuum around the H$\alpha$
line which further drops the EW in addition to making the galaxy colors
redder.  After enough time has elapsed after the burst the galaxy returns to
the same position in the color-H$\alpha$ EW plane it would have occupied had
no burst occurred, although it will have taken longer to get there.

\begin{figure}
\epsscale{0.75}
\plotone{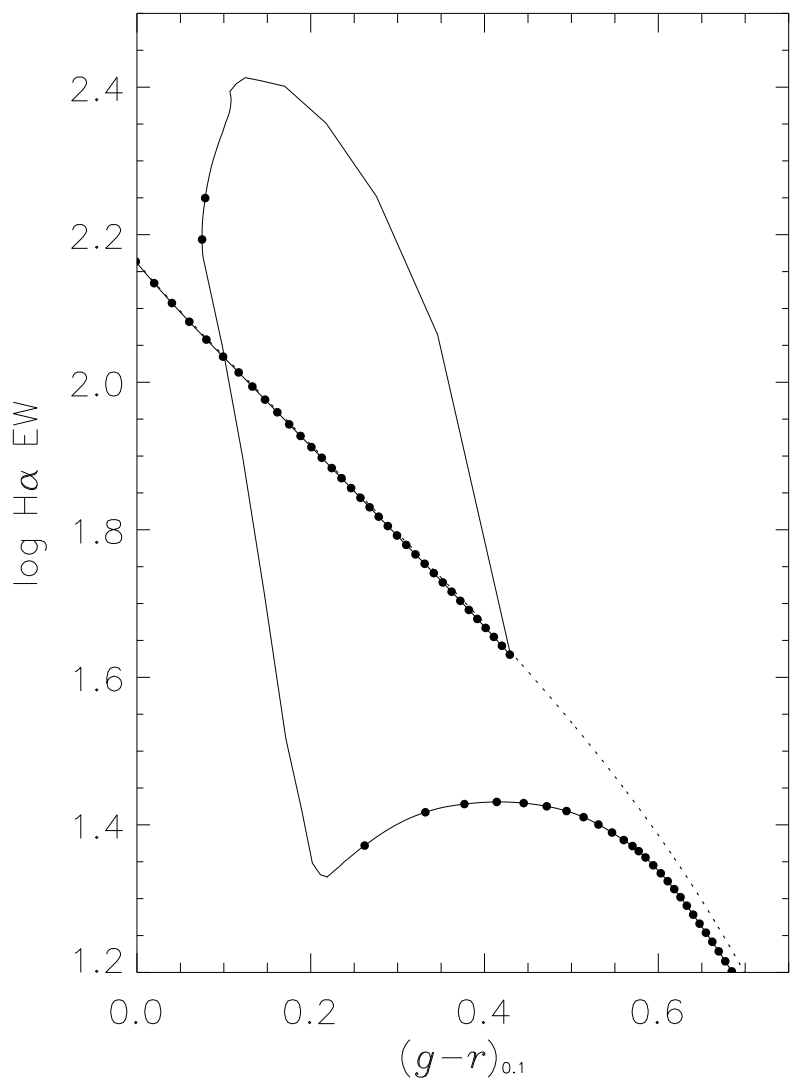}
\caption{Color and H$\alpha$ EW values as a function of age for a model galaxy experiencing a burst (solid line).  The model galaxy has solar metallicity, $\Gamma=1.35$, and an exponentially decreasing SFH with $\tau=2.15$ Gyr.  The burst occurs at an age of 4.113 Gyr, lasts 250 Myr and has a strength of 10\% of the total stellar mass.  The age of the galaxy increases from the upper left to lower right before the bursts and the black dots appear on the track at 100 Myr intervals.  The dotted line shows the track for the model had a burst not occurred.}
\label{fig:burst}
\end{figure}

Figure \ref{fig:burst} gives one example of this cycle.  A solar metallicity
galaxy with $\Gamma=1.35$ and an exponentially decreasing SFH with $\tau
=2.15$ Gyr experiences a burst of star formation at an age of 4.113 Gyr which
lasts 250 Myr and forms 10\% of the stellar mass.  The black dots, spaced at
100 Myr intervals, show that comparatively more time is spent below the
nominal track than above it.  The peak H$\alpha$ EW is reached just 5 Myr
after the start of the burst.  If you happen to be observing the galaxy during
the burst a shallower IMF will be measured (assuming a burst-free SFH), after
the burst for 1 Gyr a steeper IMF will be measured  and after that the effects
of the burst largely disappear, although the galaxy will appear younger than it
actually is.

\begin{figure}
\plotone{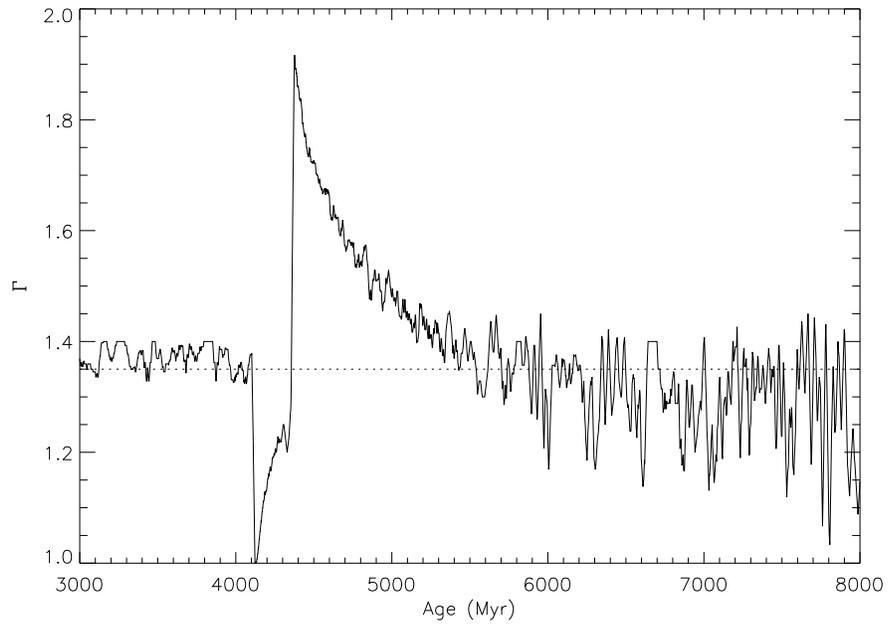}
\caption{Best fitting IMF slope as a function of age for the model galaxy in Figure \ref{fig:burst}, which experiences a burst of star formation at an age of 4.113 Gyr.  The dotted line indicates the underlying IMF model ($\Gamma=1.35$).  The $\Gamma$ values have been boxcar smoothed by 20 Myr for clarity.}
\label{fig:burstfit}
\end{figure}

Figure \ref{fig:burstfit} gives the best fitting IMF slope from our analysis,
which assumes no bursts, as a function of age for the galaxy in Figure
\ref{fig:burst}.  The jitter in the best fit $\Gamma$ values is due to the
discrete nature of our model grid, and the fact that the model tracks for
different IMFs run together at large ages.  Within a 300 Myr period during and
just after the burst the best fitting IMF slope is anywhere from $\Gamma =
1.00$ to 1.95.  This shows that even if a universal IMF exists the SFH can mimic a
huge range of IMF models.  Roughly 1 Gyr after the burst the measured IMF is
back to its true value.  A galaxy with a bursty SFH viewed at a random time
will be biased toward a steeper IMF than what is the true IMF.  While one
model is not an exhaustive study of the effects of the SFH on IMF measurements
it does give a good sense of what issues arise.

To eliminate this uncertainty we investigated cutting the sample on SFH.  In
order to detect a relative lack or excess of present star formation it is
necessary to measure both the present and past star formation rates or at
least be able to compare the two in some way.  The problem lies in the fact
that conversions of observables into star formation rates assume an IMF to do
so.  Our aim is to measure the IMF so we cannot make strong a priori
assumptions about it.  Instead of biasing or results from the start we fit all
galaxies and then try to determine the affect of SFHs on our conclusions.

\subsubsection{Single Burst Models}

The simplest burst model is that of a single burst at a random time on top of
our smooth exponential SFHs.  A grid of 1000 SFHs was constructed by first
selecting one of the 24 smoothly varying SFHs at random.  A burst lasting 200
Myr was superposed on the SFH at a time selected uniformly at random over a
range of 12.5 Gyr.  The strength of the burst was randomly selected up to 40\%
of the total stellar mass, with preference given to smaller bursts.  The
colors and EWs of these SFHs were calculated at 1 Myr intervals over 12.5 Gyr
for the IMFs $\Gamma = 1.35$ and 1.80 and a fixed metallicity of $Z=0.01$.  This metallicity choice is  based on the luminosity-metallicity relationship in \citet{CAT} for $M_{r,0.1} = -17$ galaxies.  Plotting these models in
the color-H$\alpha$ EW plane shows that all observed data points are covered
by either IMF.

To test whether the observed distribution of points in the ${\rm
  M}_{r,0.1}=-17$ bin can be explained by bursting SFHs the MC
techniques of \S 5 were used.  The models of \S 4 were replaced with
the grid of single burst models.  The 100 MC simulations were
  constructed as described earlier, but using only the 329 ${\rm
  M}_{r,0.1}=-17$ galaxies as a basis.  The analysis yielded
  $\overline{\chi^2}=0.027$ for $\Gamma=1.35$ and
  $\overline{\chi^2}=0.020$ for $\Gamma=1.80$, both of which are over  fits. 

This shows than an individual galaxy can be fit with an arbitrary IMF given the
freedom to choose a SFH.  However, our advantage is that we have \textit{many}
galaxies and the distribution of the properties of the best fit models can be
shown to be implausible.

\begin{figure}
\plotone{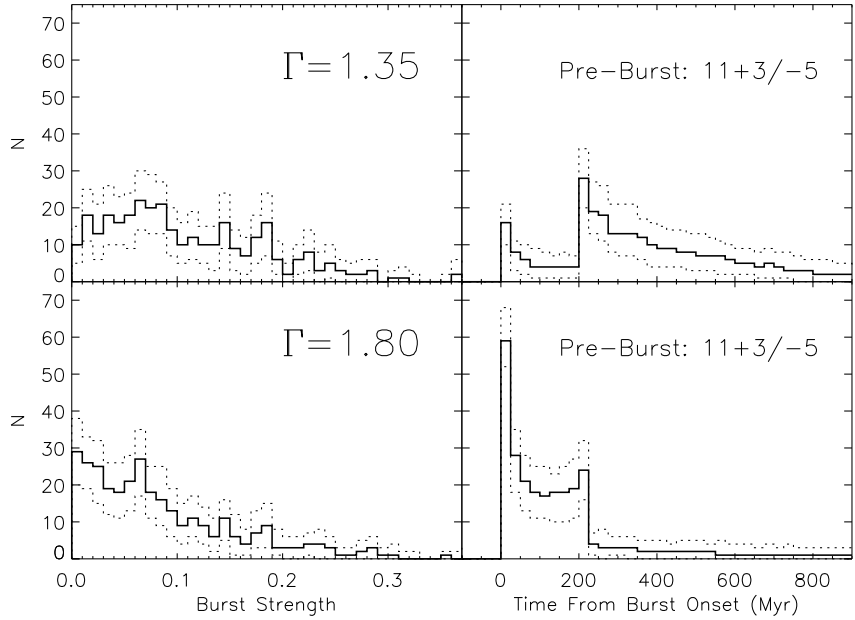}
\caption{Distributions of best fitting parameters for the synthetic ${\rm M}_{r,0.1}=-17$ data for models with $Z=0.01$ and a single burst SFH.  In all panels the dotted lines indicate the 95\% errors determined from MC simulations.  In the top
two panels are results for $\Gamma=1.35$.  In the bottom two panels are results
for $\Gamma=1.80$.  In the left column is the distribution of best
fitting burst strengths given as a fraction of the total stellar mass
formed.  At right is the distribution of best fitting ages measured
relative to the onset of the bursts.  The bursts of star formation
begin at 0 and end at 200 Myr.  For both IMFs $11+3/-5$ of the 329
galaxies are best fit by models which have yet to experience their
bursts and are not plotted.}
\label{fig:chibursts}
\end{figure}

Figure \ref{fig:chibursts} demonstrates the problem with the single burst
model.  On the left of the figure the distribution of the best fitting
burst strengths are plotted as a fraction of the total stellar mass
formed.  At right is the distribution of the best fitting times from
the burst onset.  For example, a galaxy which best fits a model with a
burst at 1.000 Gyr at an age of 1.211 Gyr has a time from burst onset
of 211 Myr.  This measure is used because in an investigation of the
effects of bursts the age relative to the burst is more important than
the age given that the bursts occur at different, random times across
the models.

For both $\Gamma=1.35$ and $\Gamma=1.80$ the number of objects best
fit by a model prior to the burst is $11 +3/-5$, or 3\%, and are not
plotted.  In both cases the distributions of the best fitting ages and
ages at which the burst begins are roughly uniform.  Given this fact
it is expected that half of the galaxies should be best fit by a
pre-burst model.  Furthermore the right hand panel shows sharp,
significant discontinuities in the distribution of best fitting time
from burst onset.  Again, viewed at random times this distribution
should be uniform but is highly peaked in the 25 Myr at the start of
the burst and the 25 Myr just after the burst ends.  In both cases the
errors bars show that the discontinuities are significant.  In the
case of $\Gamma=1.35$, 3.5 times as many galaxies are in the 200 Myr
after the burst ends than the 200 Myr during it and this 400 Myr
accounts for 57\% of all galaxies.  Although our sample is $r$-band
selected the stars in the 0.7 to 3 M$_\sun$ range which dominate the
red continuum in the red giant phase do not start to leave the main
sequence for 300 Myr.  The sharp increase in galaxies fit at 200 Myr
after the burst cannot be due to a selection effect.  Assuming a
universal IMF this points to a strong coordination of SFHs across a
population of galaxies unrelated in space.  These arguments
show that while a single burst model can fit the data extremely well,
it does not do so in a physically self-consistent fashion.

\subsubsection{Multiple Burst Models}

To find a physical motivation for SFH models for low luminosity galaxies we
look towards the Local Group.  There have been a number of recent studies of
the SFHs of local dwarf galaxies which use HST to get color-magnitude diagrams
(CMD) of resolved stellar populations.  The SFH is determined by fitting
isochrones to the CMD.  The sample here is biased by Local Group membership
and by what galaxies have been observed to date.  The galaxies mentioned here
give a point of reference rather than a well-defined distribution of SFHs.

The blue compact dwarf (BCD) UGCA 290 was found to quiescently form stars over
the past Gyr up until a ten-fold increase in SFR from 15 to 10 Myr ago which
more recently has decreased to a quarter of its peak value \citep{Crone02}.
The dwarf irregular IC 1613, which is relatively isolated and non-interacting,
was found to have SFR enhanced by a factor of 3 from 3 to 6 Gyr ago without
evidence of  strong bursts \citep{Skill03}.  The dwarf irregular NGC 6822,
also relatively isolated, is found to have a roughly constant SFH
\citep{Wyder03}.  The BCD NGC 1705 is found to be \textit{gasping}- a SFH
marked by moderate activity punctuated by short periods of decreased star
formation \citep{Annibali03}.  The authors also note that NGC 1705 is best fit
by an IMF with $\Gamma=1.6$.  NGC 1569 likely experienced three strong bursts
in the last Gyr as well as a quiescent phase from 150 to 300 Myr ago
\citep{Angeretti05}.

Inspired by the preceding Local Group SFHs we constructed six SFH classes with
multiple bursts.  These SFH classes are described in Table \ref{tab:sfhs}.
Each class starts with an underlying smooth SFH.  SFHs \#1 \& 2 have no star
formation, \#3 \& 4 have a constant SFH and \#5 \& 6 have exponentially
decreasing SFHs like those previously described.  Star formation
discontinuities are then superimposed on top of the smooth SFHs.  These
discontinuities are in the form of increased (bursts) or decreased (gasps)
star formation for periods of 200 Myr.  The time and spacing of the
discontinuities is random with the mean interval between bursts listed in Table
\ref{tab:sfhs}. 

For each SFH class described above we randomly generated 1,000 SFHs.  Colors
and EW widths were calculated for each SFH using $\Gamma=1.35$ and $Z=0.01$.
According to the SDSS mass-metallicity relationship \citep{CAT} galaxies at
${\rm M}_{r,0.1}=-17$ will on average have $Z=0.01$.

\begin{figure}
\plotone{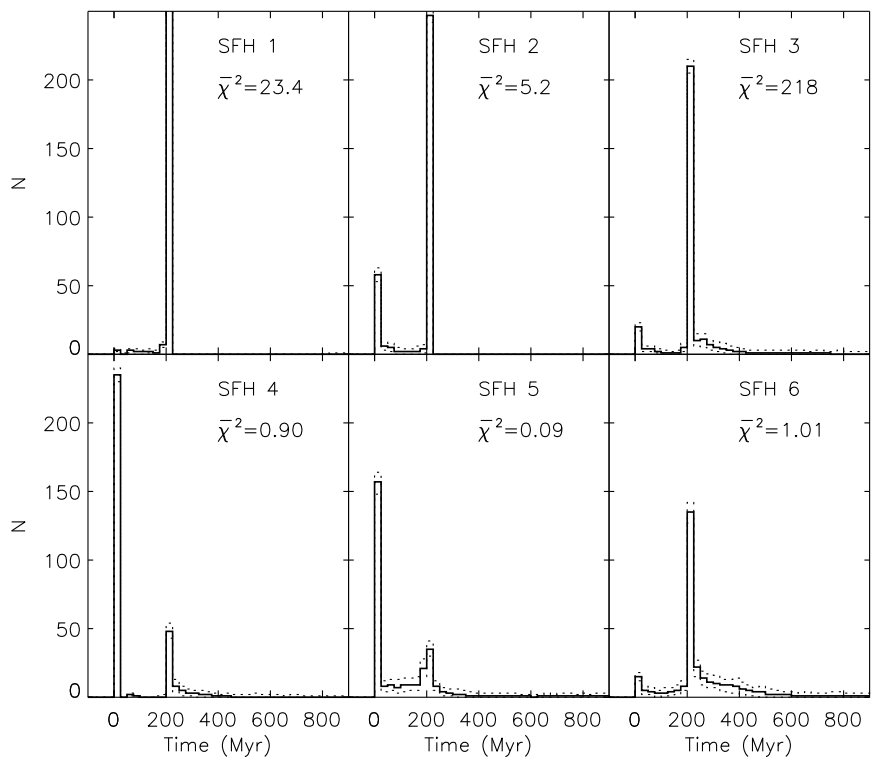}
\caption{Distributions of best fitting times measured from the
  beginning of a burst or gasp for the synthetic ${\rm M}_{r,0.1}=-17$
  data for models with $Z=0.01$, $\Gamma=1.35$ and multiple burst SFHs
  models described in Table \ref{tab:sfhs}.  The $\overline{\chi^2}$
  values for each family of models is also shown.  The dotted lines
  indicate the 95\% confidence regions determined by MC simulations.}
\label{fig:6SFH}
\end{figure}

We then repeated our $\chi^2$ analysis with the 100 MC simulations of the ${\rm
  M}_{r,0.1}=-17$ galaxies in the same manner as for the single burst models.  The results of our analysis for each of the six
  SFH classes are shown in Figure \ref{fig:6SFH}.  SFH 5, the gasps on top of
  exponential SFHs, is the best fit with $\overline{\chi^2}=0.09$.  Extended
  periods of no star formation punctuated by bursts (SFHs \# 1 \& 2) do not
  fit the data.

As was the case for the single burst models an unreasonable fraction of
galaxies are best fit by SFHs in the 20 Myr immediately following a burst or
the first 20 Myr of a gasp.  If the SFH models are reasonable we should see
roughly equal numbers of galaxies in each time bin.  There is no reason why
all of the low luminosity galaxies across the large volume of space in the
SDSS footprint should have experienced coordinated bursts.  However each panel
of Figure \ref{fig:6SFH} has at least 40\% of the galaxies in one 20 Myr bin.

One explanation for this is that it is an artifact of our sample being
selected in the $r$ band.  However spectral synthesis models show that for
instantaneous bursts of star formation the $r_{0.1}$ magnitude is brightest at
the burst time and decays smoothly for a range of $\Gamma$.  If anything it is
more likely to catch galaxies during a burst rather than after or after a gasp
instead of during one.

Regardless of the SFH model the presence of blue galaxies with low H$\alpha$
EW requires a recent discontinuity in the SFR for $\Gamma=1.35$.  Based on
the evolution of the $r_{0.1}$ band luminosity we expect to see a similar
number of galaxies with excess H$\alpha$ EWs.  The fact that these galaxies
are missing shows that the discontinuous SFH models do not match our
observations.  Therefore IMF variations are a more likely explanation for the
observed distribution of ${\rm M}_{r,0.1}=-17$ galaxies.

\subsubsection{Recovering $\Gamma$ from Synthetic Data}

As a last exercise the best fitting SFH models from the previous section can be run forward to see if
the correct IMF can be recovered.  For each SFH model grids 10,000 data points
were chosen by selecting a random SFH and a \textit{uniformly distributed} age.  Normal
errors were added using the error characteristics of the ${\rm M}_{r,0.1}=-17$
bin.  This synthetic data was analyzed in the same fashion as the real data in
the earlier sections.  For the single burst models the recovered IMF models
for $\Gamma =1.35$ and 1.80 were 1.34 and 1.79 respectively with best fitting
$\overline{\chi^2}=$ 0.80 and 0.65.  For SFH 5 the recovered IMF was also
$1.34$ with $\overline{\chi^2}=1.5$.  In all three cases the correct IMFs were
recovered although the fit was worsened by by the burst activity.  This
reinforces the difficulty in producing enough blue, low H$\alpha$ EW galaxies
to match the observed data with simple SFH models.

\section{H$\delta_{\rm A}$ Absorption}

In the previous sections we have expanded on the K83 method and exploited the
H$\alpha$ and color information as much as possible.  In the bias section we
found that various possible biases do not fully explain either the increased
values of $\Gamma$ or the poor fit to a single IMF in the lowest luminosity
bins.  In the SFH section we found that an arbitrary $\Gamma$ value coupled
with a plausible SFH with bursts or gasps can account for the position of any
individual galaxy in the color-H$\alpha$ EW plane.  However taking the
population of ${\rm M}_{r,0.1}=-17$ galaxies together necessitates an
incredibly unlikely coordination of SFHs across the disparate group of
objects.  This points to the extraordinary conclusion that while the IMF may
be universal across luminous galaxies, it is not in fact universal in low
luminosity galaxies.  Such an extraordinary claim would ideally be backed by
extraordinary evidence.  In this section we take a look beyond the K83 method
for some reinforcement of our result.

The H$\delta$ absorption feature can be used to gain additional insight into
the nature of stellar populations.  H$\delta$ absorption is due to absorption
lines form stellar photospheres.  The Balmer absorption lines are most
prominent in A stars and weaken due to the Saha equation for both hotter and
cooler stars.  As such the H$\delta$ absorption is a proxy for the fraction of
light of a stellar population being supplied by A stars, and to a lesser
extent B and F type stars.  In a stellar population of a uniform age the
H$\delta$ absorption will peak after the O and B stars burn out, but before
the A stars leave the main sequence.  For this reason the strength of
H$\delta$ can be used help determine the age of a population or to detect
bursts of star formation which occurred around 1 Gyr in the past.

\citet{Worthey97} describes two different methods for measuring H$\delta$
absorption.  The H$\delta_{\rm F}$ definition is tuned to most accurately
measure the H$\delta$ absorption from F stars.  The H$\delta_{\rm A}$
definition has a wider central bandpass to match the line profiles of A stars.
They state that the H$\delta_{\rm A}$ definition is better to use for galaxies
because it is less noisy in low S/N galaxies and velocity dispersion acts to
widen absorption features.  On the downside the narrower H$\delta_{\rm F}$
definition is much more sensitive to population age where it can be used.
\citet{Worthey97} observationally determines the range of H$\delta_{\rm A}$
values to be from 13 for A4 dwarf stars to -9 for M-type giant stars.

H$\delta_{\rm A}$ values can be measured from the SDSS spectra.  Like the
H$\alpha$ EW values our H$\delta_{\rm A}$ values come from \citet{CAT} instead
of the SDSS pipeline.

\begin{figure}
\plotone{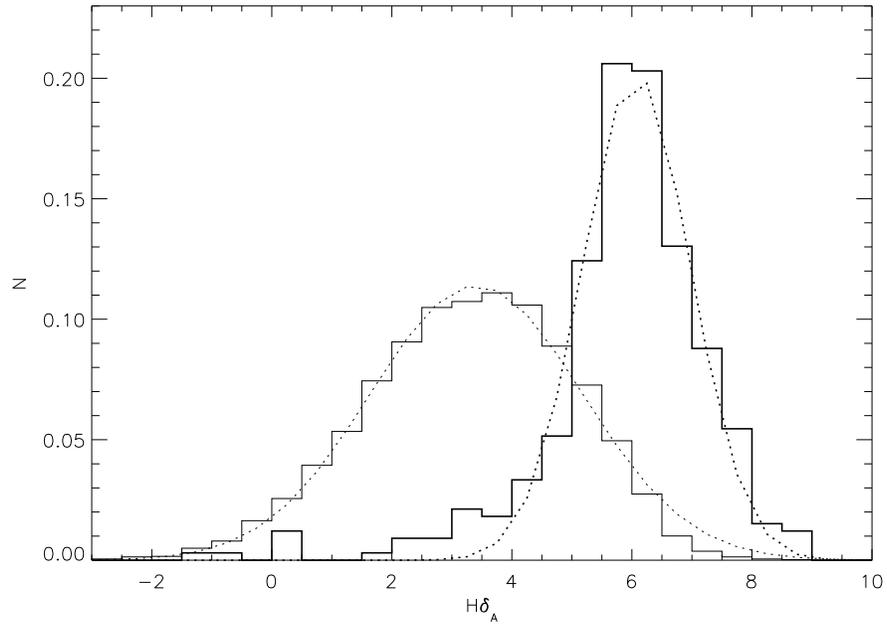}
\caption{Histogram of the measured H$\delta_{\rm A}$ values for the ${\rm
    M}_{r,0.1}=-17$ (bold line) and ${\rm M}_{r,0.1}=-23$ (thin line) volume
  limited bins from Figure \ref{fig:hagr1723}  expressed as a fraction of the
  total number of galaxies in each luminosity bin.  The median uncertainty in
  H$\delta_{\rm A}$ is 0.9 in both bins.  The dotted lines are Gaussian
  profiles fitted to the distributions.  For the ${\rm M}_{r,0.1}=-23$ bin
  (thin dots) the profile is centered on 3.4 with $\sigma = 1.8$ and for ${\rm
  M}_{r,0.1}=-17$ bin (thick dots) the profile is centered on 6.1 with
  $\sigma = 0.9$.}
\label{fig:hdfig}
\end{figure}

Figure \ref{fig:hdfig} compares the distribution of the H$\delta_{\rm A}$
values for the ${\rm M}_{r,0.1}=-17$ and ${\rm M}_{r,0.1}=-23$ luminosity
bins.  For reference recall that Figure \ref{fig:hagr1723} plots the color and
H$\alpha$ EWs for these two bins.  The difference between the high and low
luminosity bins is clear.  Gaussian profiles can be fit to both distributions.  For
the ${\rm M}_{r,0.1}=-23$ bin the standard deviation of the profile is twice
as large as the measurement error in H$\delta_{\rm A}$ suggesting that the
true distribution has a range of values.  Assuming both the errors and
underlying distribution are Gaussian the distribution of H$\delta_{\rm A}$ for
the luminous galaxies is centered at H$\delta_{\rm A} =3.4$ with a standard deviation of 1.5.  By contrast, the Gaussian fit to the ${\rm M}_{r,0.1}=-17$ bin has the same standard deviation
as the median error in the galaxies.  This is consistent with nearly all of
the galaxies having H$\delta_{\rm A} =6.1$. 

This shows that the fainter galaxies on average have significantly larger
fractions of A-type stars amongst their stellar populations.  It also shows a
seemingly unlikely coordination of H$\delta_{\rm A}$ in the low luminosity
galaxies, reminiscent of the earlier suggestion of coordinated SFHs.  The
question then becomes why? 

For a possible explanation we look again to the models.  Modeling the behavior
of H$\delta_{\rm A}$ requires an extra step.  The standard PEGASE.2 models do
not have the required resolution to accurately measure the H$\delta_{\rm A}$
index.  This is remedied with the use of the PEGASE-HR code \citep{LeBorgne}
which uses a library of echelle spectra of 1503 stars to calculate spectral
synthesis models with R=10,000 over the range of 4000 to 6800\AA.  Using
PEGASE-HR in the low resolution mode yields the same results as PEGASE.2, and
the same input parameters are used for both codes.  The models here are the
same as those described earlier in the models section, but have been
recalculated using PEGASE-HR to allow H$\delta_{\rm A}$ measurements.

\begin{figure}
\plotone{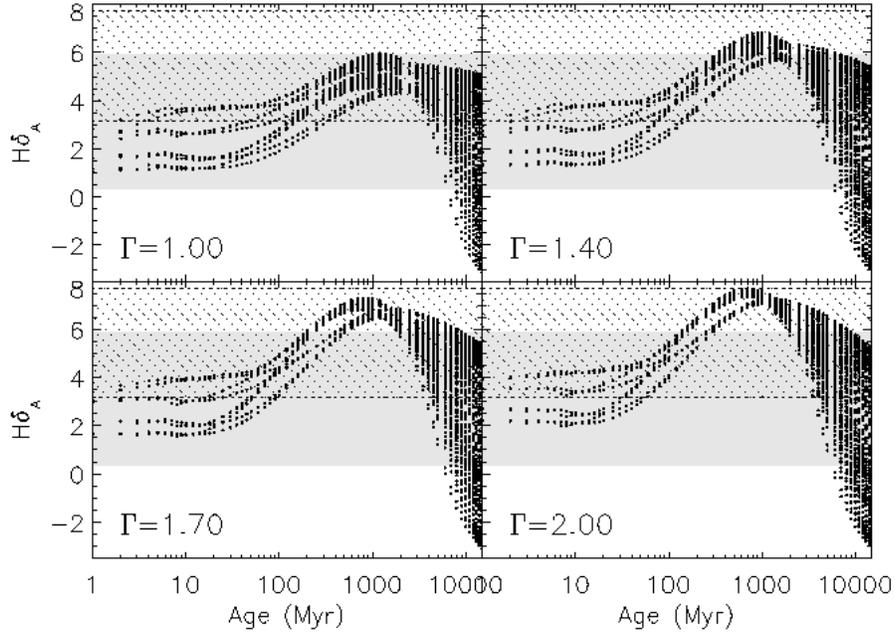}
\caption{Model values of the H$\delta_{\rm A}$ index as a function of age for
  four IMF models, $\Gamma=1.00$ (upper left), 1.40 (upper right), 1.70 (lower
  left), and 2.00 (lower right).  Each dot represents a model value for single
  IMF, metallicity, SFH (as described in the models section) and age.  The
  four tracks identifiable towards the left of each panel are each of
  differing metallicities with the lowest metallicity models having the
  largest H$\delta_{\rm A}$ values at young ages.  The grey area indicates the
  range of H$\delta_{\rm A}$ spanned by the middle 90\% of the volume limited
  ${\rm M}_{r,0.1}=-23$ bin.  The hashed area shows the middle 90\% range for
  the volume limited ${\rm M}_{r,0.1}=-17$ bin.}
\label{fig:hdmodplot}
\end{figure}

Figure \ref{fig:hdmodplot} shows the behavior of the H$\delta_{\rm A}$ index
as a function of age for four different IMF models, $\Gamma=1.00$, 1.40, 1.70
and 2.00.  For each IMF models of all metallicities and smooth SFHs are
plotted for each age.  The qualitative behavior of all models is the same.
The H$\delta_{\rm A}$ holds steady for the first 20 to 40 Myr before
increasing to a peak value at 700 Myr to 1 Gyr and then falls off.  Prior to
reaching the peak value for each IMF the metallicity has the strongest effect
on the H$\delta_{\rm A}$ index.  At this point the lowest metallicity galaxies
have the highest H$\delta_{\rm A}$.  After the peak the SFH has the strongest
effect with the constant and increasing SFHs maintaining higher H$\delta_{\rm
  A}$  values.

After a few Gyr the differences in H$\delta_{\rm A}$ between models with
different IMFs disappear.  Prior to that there are three main differences.
First, the peak H$\delta_{\rm A}$ values are higher for larger values of
$\Gamma$.  For $\Gamma=2.00$ H$\delta_{\rm A}$ reaches a maximum value of
nearly 8 and the maximum value is similar for all metallicities.  For
$\Gamma=1.00$ the maximum value ranges from 4.5 to 6 depending on the
metallicity.  This is due to the fact that the steeper IMFs have fewer
luminous massive stars to dilute the H$\delta$ absorption features from the A
star population.  Secondly, the low $\Gamma$ models have H$\delta_{\rm A}$
values that start their initial increases at a later time.  Lastly the low
$\Gamma$ models reach their peak values later in time than those with fewer
massive stars.

To compare the H$\delta_{\rm A}$ values for the ${\rm M}_{r,0.1}=-23$ and
$-17$ bins to the models the range of the middle 90\% values from Figure
\ref{fig:hdfig} for each bin are overlaid on Figure \ref{fig:hdmodplot}.  Once
again the ${\rm M}_{r,0.1}=-23$ bin is in good agreement with our assumption
of a universal IMF and smooth SFH.  The range of H$\delta_{\rm A}$ values can
be accomplished with a single Salpeter-like $\Gamma=1.40$ IMF with
only the proviso that most galaxies be older than a few Gyr or younger
than 300 Myr.  However the exact same statement can be made for $1.0 < \Gamma < 2.0$ so H$\delta_{\rm A}$ provides only a constraint on the age of the most luminous galaxies, but not the IMF.

For the ${\rm M}_{r,0.1}=-17$ galaxies the distribution of H$\delta_{\rm A}$
cannot be achieved with the shallowest IMFs investigated under the assumption
of smooth SFHs.  However Salpeter and steeper IMFs can be
accomplished.  What changes is the range of ages over which the models
have the correct H$\delta_{\rm A}$.  Steeper IMFs require that the
galaxies be either older than a few Gyr or 200 Myr old to accommodate
the observed H$\delta_{\rm A}$ values.  Most troubling is that there
does not appear to be any reason why H$\delta_{\rm A}$ should stack up
at 6.1 for the ${\rm M}_{r,0.1}=-17$ galaxies. 

H$\delta_{\rm A}$  values can also be calculated for the same SFH class models used earlier.  Unfortunately this does not provide any added constraints.  H$\delta_{\rm A}$  values remain elevated for several 100 Myr after the start of a burst or the end of a gasp of SFH.  The behavior of H$\delta_{\rm A}$ in the presence of SFH discontinuities provides no need for galaxies to stack up in the narrow 20 Myr intervals seen in the earlier multiple burst section, nor do these models suggest why the low luminosity galaxies are consistent with a single value of H$\delta_{\rm A}$.

There is no satisfactory model to account for the H$\delta_{\rm A}$ distribution of the low luminosity galaxies.  However the SFH results from the previous section strongly suggest that the incredible coordination of discontinuities is highly unlikely.  Steeper IMFs do allow for higher H$\delta_{\rm A}$ values for longer periods of time, thus relaxing the SFH coordination requirement.  This agrees with our earlier results for faint galaxies.  The low luminosity galaxies are most likely the result of a mix of IMFs which are on average steeper than Salpeter.

\section{Conclusions}

The goal of this paper was to revisit the K83 method for inferring the IMF
from integrated stellar populations and to harness the richness of the SDSS
data, improved spectral synthesis models and greater computational power
available today to make a state-of-the-art measurement of the IMF.  The
quality of the SDSS spectroscopy allowed us to address several of the
limitations of K83 and KTC94-- we resolve the [\ion{N}{2}] lines (a
significant improvement in the accuracy of the H$\alpha$ EWs of individual
galaxies), eliminate contamination from AGN, make extinction corrections for
individual galaxies and fit underlying stellar absorption of H$\alpha$.  We
succeeded in achieving more accurate EWs for individual galaxies.  The median
total EW error for our sample is 17\% compared to a 10\% uncertainty in EWs
combined with a 20-30\% uncertainty in the extinction correction for K83.

We expanded the grid of models to allow for a range of ages and metallicities.
We used $\chi^2$ minimization to go beyond differentiating between two or
three IMF models to actual fitting for the best IMF slope.  The vast size of
the SDSS sample allowed us to both drive down random errors and to cut
the data into narrow parameter ranges which were still statistically viable.

The size of the DR4 sample yielded $\Delta \Gamma=0.0011$ 95\% confidence
region due to random error for the sample as a whole.  Even the volume limited
${\rm M}_{r,0.1}=-17$ luminosity bin with only 329 objects has a random error
of $\Delta \Gamma=0.0086$.  Only in bins with fewer than 10 objects do the
random errors become significant.  Our IMF fitting is therefore dominated by
systematics.

Originally we believed that our systematics would be dominated by the effects
of SFH discontinuities.  However we conducted several experiments where we
selected populations of galaxies from models with bursting or gasping SFHs and
gave them measurement errors consistent with those in the ${\rm
  M}_{r,0.1}=-17$ luminosity bin.  To our surprise our $\chi^2$ minimization
revealed the true IMFs with $\Delta \Gamma=0.01$.  The main effect was to
reduce the quality of the fits.  This is due to the fact that H$\alpha$ EWs
return to nominal levels in a relatively short time after SFH
discontinuities.

\begin{figure}
\plotone{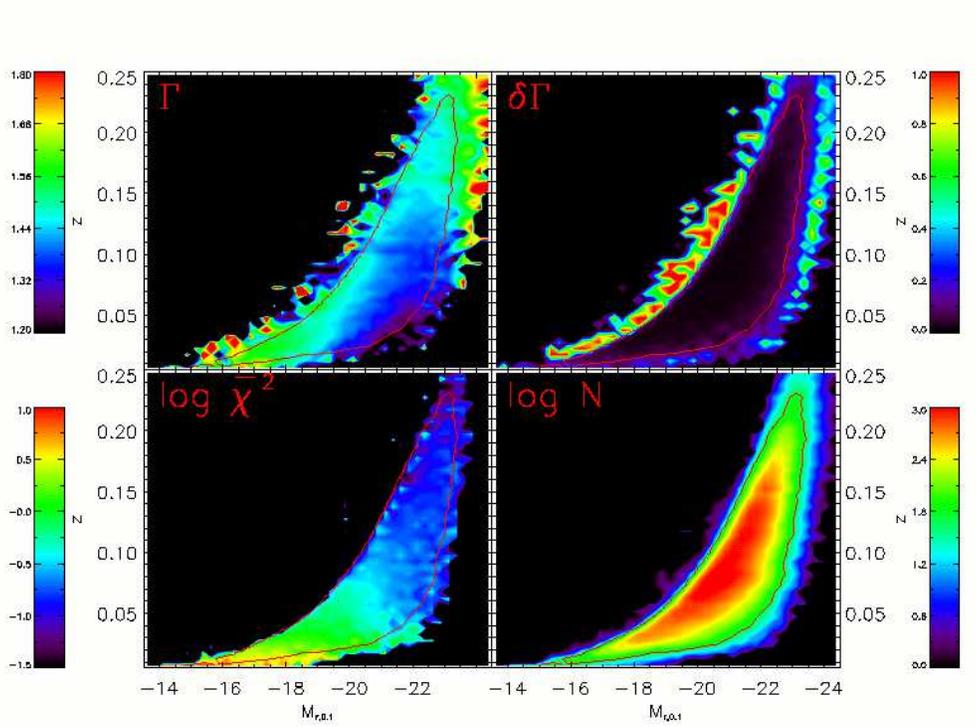}
\caption{Values of fitted parameters as a function of both
  luminosity and redshift.  The bins extend 0.25 magnitudes in luminosity and
  0.005 in redshift.  Clockwise from upper left the frames show the best
  fitting $\Gamma$, the width of the 95\% confidence region in $\Gamma$ from
  the MC simulation, the log of the number of galaxies in each two dimensional
  bin and the log of $\overline{\chi^2}$.  The shading levels for each panel
  are given by the adjacent vertical color bars.  Black areas indicate regions
  where there are no galaxies with the respective combination of redshift and
  luminosity.  The red contour indicates the region in which there are at
  least 50 galaxies in each 2-D bin.}
\label{fig:redshift}
\end{figure}

\begin{figure}
\plotone{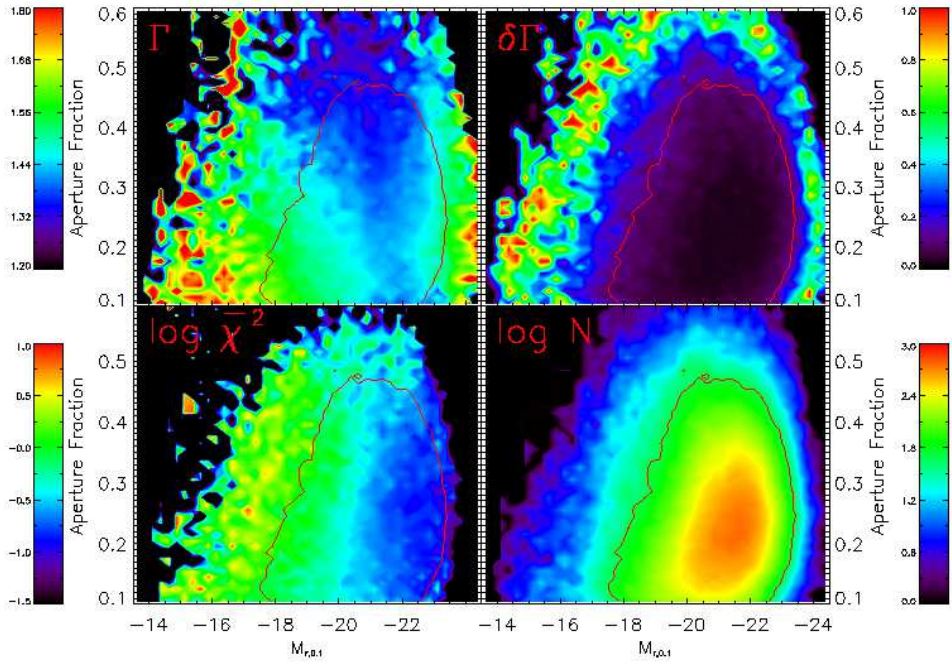}
\caption{Values of fitted parameters for 2-D bins cut on luminosity and
  aperture fraction.  The bins extend 0.25 magnitudes in luminosity and
  1\% in aperture fraction.  The description is identical to Figure
  \ref{fig:redshift} as are the shading levels.}
\label{fig:aperture}
\end{figure}

\begin{figure}
\plotone{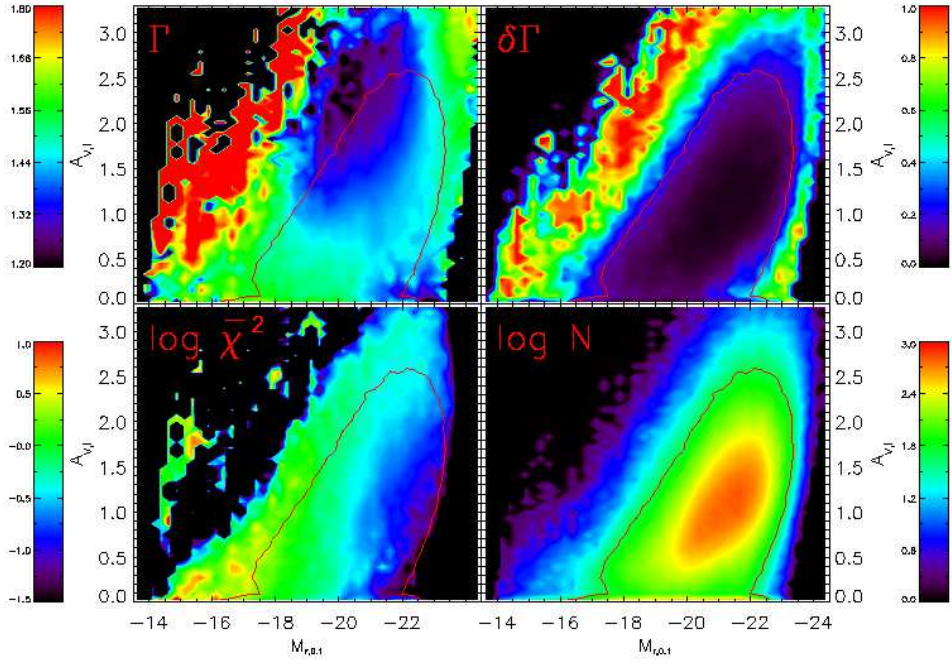}
\caption{Values of fitted parameters for 2-D bins cut on luminosity and
  emission line extinction.  The bins extend 0.25 magnitudes in luminosity and
  0.06 magnitudes in extinction.  The description is identical to Figures
  \ref{fig:redshift} and \ref{fig:aperture}, as are the shading levels.}
\label{fig:extinction}
\end{figure}

Another way to estimate the size of the systematic errors is to look at the
trends of the ${\rm M}_{r,0.1}=-21$ and $-22$ luminosity bins, because they
have the largest membership, in Figures \ref{fig:redshift}, \ref{fig:aperture}
and \ref{fig:extinction}.  Assuming that the IMF is universal and that our
method is perfect we should get the same answer for any subset of the data we
might choose.  The largest ranges are $\Delta \Gamma \sim 0.12$ for redshift
binning, $\Delta \Gamma \sim 0.19$ by aperture and $\Delta \Gamma \sim 0.19$
by extinction.  Conservatively then the systematic error is $\pm 0.1$.

\begin{figure}
\plotone{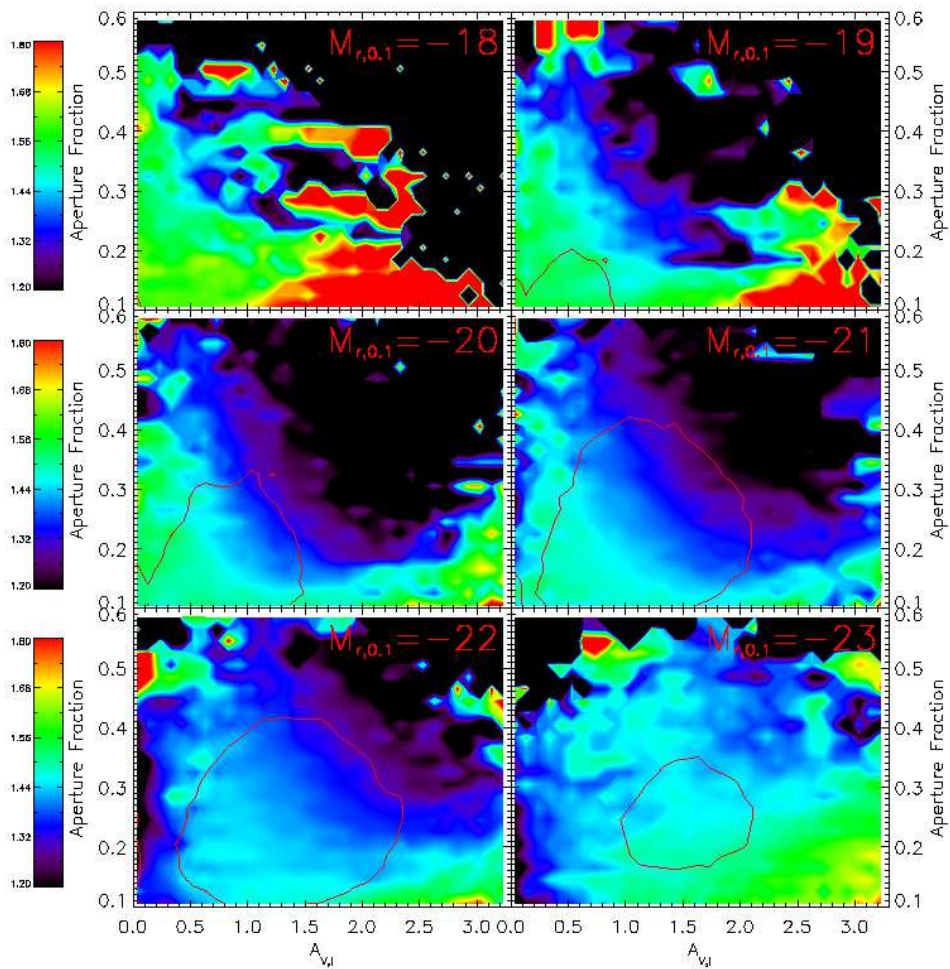}
\caption{Best fitting $\Gamma$ values as a function of aperture fraction and
  measured emission line extinction for six volume limited luminosity bins.
  The 2-D bins extend 0.1 magnitudes in extinction and 2\% in aperture
  fraction.  The volume limited bins are defined as described in Figure
  \ref{fig:VL-plot}.  The red contour shows the area where the 2-D bins
  contain at least 50 galaxies.  The shading levels are described by the color
  bars on the left.}
\label{fig:avap}
\end{figure}

There are two points to be kept in mind about this estimate of the systematic
error.  For one it is the systematic error in the exact value of $\Gamma$.
Even in Figure \ref{fig:avap} where more narrow bands of measured
extinction and aperture fraction are considered the same trends with
luminosity are seen as with the sample as a whole.  The relative systematics
between luminosity bins in these narrow slices is much smaller.  The second
thing to remember is the way in which we empirically defined our systematic
error discounts the possibility of IMF variations.  What we have called
systematics could actually be science.  If galaxies have radial IMF gradients
or if dust content plays a strong role in star formation the systematic error
could be much smaller.  The main area in which we were unable to improve upon
the K83 and KTC94 studies is that they were able to match the aperture size to
the galaxies which avoids the issue of aperture effects.

In spite of a more quantitative approach, like K83 and KTC94 the results are
mostly qualitative.  However there are four key results from our
investigation.

First, for galaxies brighter than ${\rm M}_{r,0.1}\sim-20$ the best fitting
IMFs are Salpeter-like ($\Gamma \sim 1.4$).  In addition the assumption of a
universal IMF and smoothly varying SFHs is a good fit.  This is reassuring as
it follows the conventional wisdom and provides confidence that the method
works.

Secondly, galaxies fainter than ${\rm M}_{r,0.1}\sim-20$ are best fit by
steeper IMFs with larger fractions of low mass stars.  For these galaxies a
universal IMF and smooth SFH is a poor assumption.  This result is in
qualitative agreement with evidence that LSBs have bottom-heavy IMFs
\citep{Lee04}.

Thirdly, while breaking the IMF-SFH degeneracy for individual galaxies using
the H$\alpha$ EW and color is hopeless, for a statistical sample of galaxies
the degeneracy can be broken.

Lastly, given our analysis of discontinuous SFHs it appears that the IMF is
not universal in low luminosity galaxies and fewer massive stars are being
created in these galaxies.

\begin{figure}
\plotone{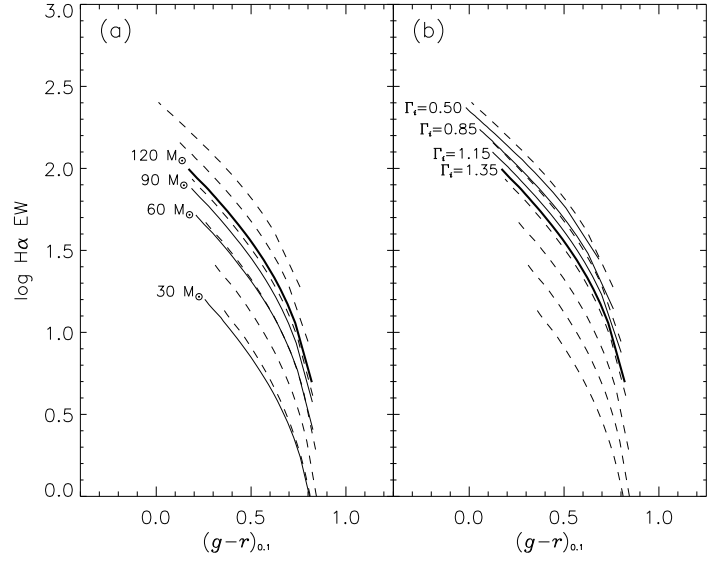}
\caption{Examples of the degeneracy of IMF parameters.  All models shown have solar metallicity, age 6 Gyr and the tracks are lines of varying SFH from a rapidly decreasing SFR on the lower right to slowly increasing on the upper left.  In both panels the bold track has $\Gamma = 1.35$ and the dashed lines have IMFs $\Gamma = $ 1.00, 1.20, 1.40, 1.60, 1.80 and 2.00 from top to bottom, all with an upper mass cutoff of 120 M$_\sun$.  (a) The solid lines are models where the upper mass cutoff of the IMF is reduced from 120 M$_\sun$ to 90, 60 and 30 M$_\sun$ as labeled.  (b) The nominal two-piece IMF is replaced by a three piece IMF in the solid lines.  In all solid tracks $\Gamma =1.35$ above 10 M$_\sun$ and $\Gamma = 0.50$ below 0.5 M$_\sun$, but is altered in the intermediate mass region from 0.5 to 10 M$_\sun$.  The solid lines have $\Gamma_i$ = 0.50, 0.85 and 1.15 over 0.5 to 10 M$_\sun$ as labeled.}
\label{fig:oddimf}
\end{figure}

\begin{figure}
\plotone{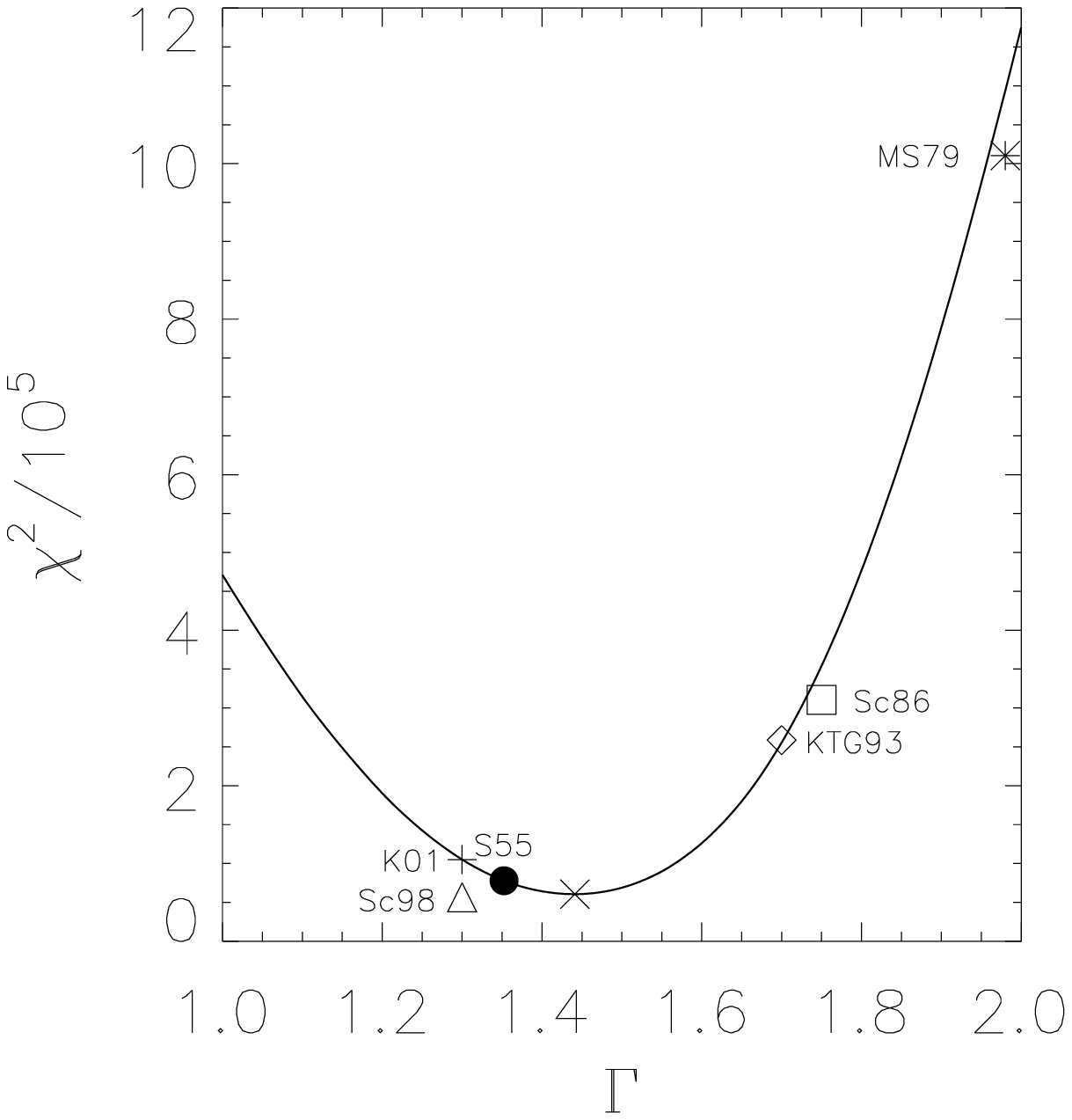}
\caption{Value of $\chi^2$ as a function of $\Gamma$ for the entire sample (solid line).  The ``X'' marks the minimum $\chi^2$ value of 61671.4 at $\Gamma = 1.4525$.  Also plotted are the $\chi^2$ for some ``classic'' IMF models, \citet{S55} (filled circle), \citet{MS79} (asterisk), \citet{Scalo86} (square), \citet{Kroupa93} (diamond), \citet{Scalo98} (triangle) and \citet{Kroupa01} (plus sign) plotted at rough estimates for equivalent $\Gamma$ values using our parameterization.}
\label{fig:all-like}
\end{figure}

\begin{figure}
\plotone{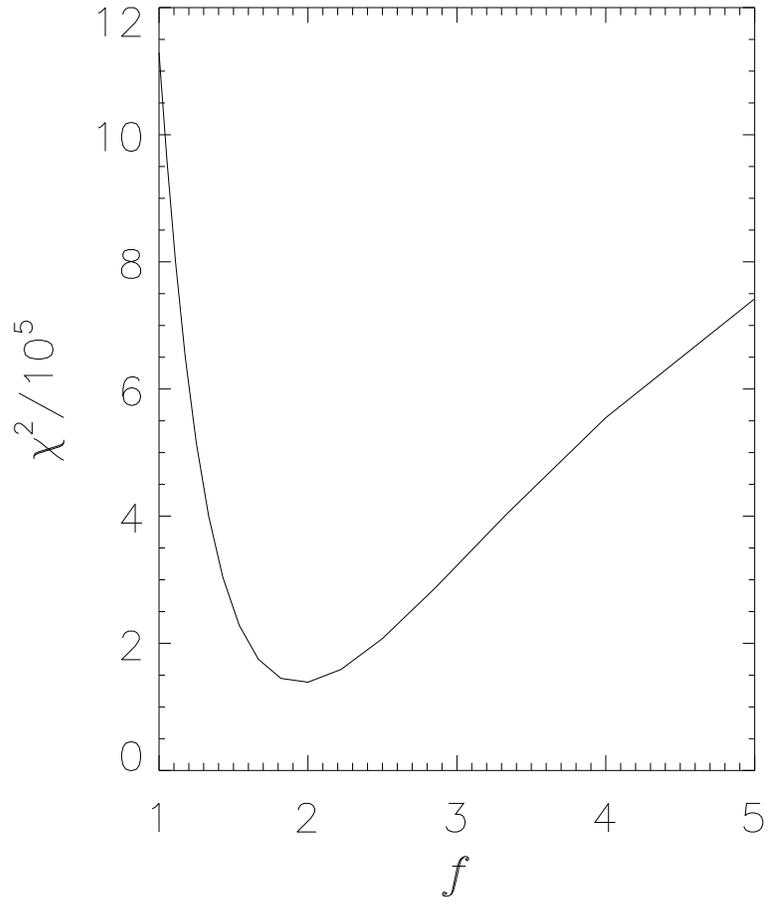}
\caption{Similar to Figure \ref{fig:all-like}, this figure shows the value of $\chi^2$ as a function of the assumed $f$ ratio for the entire sample assuming $\Gamma = 1.35$.}
\label{fig:f-all}
\end{figure}

It is worth mentioning the main caveat of our $\Gamma$ values again.  As
illustrated in Figure \ref{fig:oddimf} IMF parameterizations are themselves
degenerate in our parameter space.  Increasing the IMF slope has a similar
effect to lowering the highest mass stars that are formed or increasing the
fraction of intermediate mass stars.  This method cannot explicitly determine
if two populations have the same underlying IMF.  Figure \ref{fig:all-like}
shows that for the sample as a whole the \citet{Scalo98} three part power law
yields nearly the same result as our two part power law.  However our method
is sensitive in many cases if the IMFs are different.

In terms of star forming cloud temperatures the harsher ambient radiation and
larger number of sources of cosmic rays present in more luminous galaxies
agree qualitatively with our results.  With the extra energy hitting the star
forming clouds larger masses may be needed for contraction and fractionization
may end sooner, suppressing the formation of less massive stars
\citep{Larson98}.  \citet{CCT05} find that while the \ion{H}{2} regions of the
luminous grand design spiral NGC 5457 (M 31) can be reproduced by a single
Salpeter IMF, for the low luminosity flocculent galaxy NGC 4395 a blend of two
IMFs is required.  However, such trends are not seen in studies of
well-resolved stellar populations \citep{Kroupa02}.

Another explanation for the absence of massive stars is that the massive stars
are there, but are not visible.  Extinction to the center of star forming
regions, where massive stars preferentially exist, can reach ${\rm A_V}\sim
20$ \citep{Engelbracht98}.  However the low luminosity galaxies have the
lowest observed extinctions (see Table \ref{tab:lumbins}) which is the
opposite of what would be expected given our IMF results.

It is also possible that the IMF is in fact universal, but the way in which it
is sampled in embedded star clusters leads to an integrated galaxial IMF which
varies from the true IMF.  \citet{WK05} use a universal IMF with the
assumption that stars are born in clusters where the maximum cluster mass is
related to the star formation rate.  For a range of models this leads to a
narrow range for the apparent IMF in high mass galaxies.  For low mass
galaxies the IMF is steeper with a wider range of slopes.  The results here
are in qualitative agreement for some of the integrated galactic IMF scenarios
in  \citet{WK05} given that there should be a rough correlation between galaxy
luminosity and mass.  Once again, \citet{Elmegreen06} argues that the galaxy
wide IMF should be the same as the IMF in clusters regardless. 

In light of the theory of \citet{WK05}, whether the results of this
paper speak to a relationship between environment and the formation of
individual stars is open to interpretation.  However the impact on the
modeling and interpretation of the properties of galaxies is clear.
\citet{KWK07} note that the integrated galaxial IMF is the correct IMF
to use when studying global properties of galaxies.  Even if the IMF
of stars is in truth universal it may currently be misused in the
modeling of galaxies.  Furthermore a varying integrated galaxial IMF
could open the door to new insights in galaxy evolution.  For
instance, \citet{KWK07} suggest that the observed mass-metallicity
relationship in galaxies naturally arises from a variable integrated
galaxial IMF similar to the results of this paper.

Future work will expand in several directions.  The success constraining the
IMF with only two observed parameters (albeit carefully chosen to be
orthogonal to systematic errors) motivates a more expansive analysis with more
parameters.  Information from wavelength regimes beyond the SDSS can be used.
For instance the absorption strength and P-Cygni profile shape of the
\ion{C}{4} $\lambda 1550$ line due to massive stars is sensitive to the IMF
slope and upper mass cutoff \citep{LRH95}.  However it could be contaminated
by absorption from the interstellar medium and would need to be disentangled
\citep{Shapley03}.  A full Markov Chain Monte Carlo analysis could be
implemented fitting to multiple spectral features and marginalizing over a
range of SFHs.  In addition to luminosity, surface brightness and gas phase
metallicity can be tested for systematic IMF variations.  While the results
for luminous galaxies are already dominated by systematics, the continued
progress of the SDSS can provide better statistics for a more detailed
analysis of what physical processes are behind the IMF variations in faint
galaxies.

\acknowledgments
  
E.A.H. and K.G. acknowledge generous funding from the David and Lucile Packard
Foundation.  We would also like to thank C. Tremonti, G. Kauffmann and
T. Heckman for an early look at their SDSS spectral line catalogs.
E.A.H. would like to thank Johns Hopkins for funding from various sources as
well as C. Tremonti, E. Peng and A. Pope for invaluable assistance with the
nuances of the SDSS.  Several plots were created with the use of publicly
available IDL routines written by D. Schlegel. 

Funding for the creation and distribution of the SDSS Archive has been
provided by the Alfred P. Sloan Foundation, the Participating Institutions,
the National Aeronautics and Space Administration, the National Science
Foundation, the U.S. Department of Energy, the Japanese Monbukagakusho, and
the Max Planck Society. The SDSS Web site is http://www.sdss.org/.

The Participating Institutions are The University of Chicago, Fermilab, the
Institute for Advanced Study, the Japan Participation Group, The Johns Hopkins
University, the Max-Planck-Institute for Astronomy (MPIA), the
Max-Planck-Institute for Astrophysics (MPA), New Mexico State University,
Princeton University, the United States Naval Observatory, and the University
of Washington.

\clearpage

\begin{deluxetable}{crcccccccrcc}
\tabletypesize{\scriptsize}
\tablecaption{Luminosity Bin Details}
\tablewidth{0pt}
\tablehead{
\multicolumn{1}{c}{} & \multicolumn{8}{c}{Volume Limited} &
\multicolumn{3}{c}{All} \\
\colhead{${\rm M}_{r,0.1}$} & \colhead{$n$} & \colhead{$\Gamma_{\rm low}$} &
\colhead{$\Gamma_{\rm best}$} & \colhead{$\Gamma_{\rm high}$} &
\colhead{$\overline{\chi^2}$} & \colhead{$\overline{z}$} & \colhead{$\overline{{\rm A}_{{\rm V},l}}$} &
\colhead{$\overline{\rm ap}$} & \colhead{$n$} & \colhead{$\Gamma_{\rm best}$} &
\colhead{$\overline{\chi^2}$}}
\startdata
-14 & -       & -     & -     & -     & -    & -     & -    & -    & 28     & 1.3835 & 7.43 \\
-15 & -       & -     & -     & -     & -    & -     & -    & -    & 188    & 1.3892 & 9.86 \\
-16 & -       & -     & -     & -     & -    & -     & -    & -    & 406    & 1.5461 & 4.64 \\
-17 & 329     & 1.6000 & 1.6045 & 1.6086 & 3.41 & 0.013 & 0.38 & 0.20 & 1,304  & 1.5879 & 2.57 \\
-18 & 1,555   & 1.5338 & 1.5370 & 1.5424 & 2.05 & 0.022 & 0.50 & 0.21 & 4,327  & 1.5326 & 1.57 \\
-19 & 4,935   & 1.4772 & 1.4788 & 1.4813 & 1.17 & 0.032 & 0.70 & 0.21 & 10,375 & 1.4813 & 0.96 \\
-20 & 12,951  & 1.4306 & 1.4320 & 1.4330 & 0.63 & 0.050 & 0.91 & 0.22 & 24,851 & 1.4436 & 0.56 \\
-21 & 28,633  & 1.4051 & 1.4057 & 1.4063 & 0.32 & 0.077 & 1.15 & 0.24 & 41,411 & 1.4064 & 0.31 \\
-22 & 29,701  & 1.4036 & 1.4042 & 1.4050 & 0.19 & 0.116 & 1.32 & 0.25 & 38,406 & 1.4084 & 0.20 \\
-23 & 8,049   & 1.4545 & 1.4556 & 1.4568 & 0.15 & 0.168 & 1.55 & 0.27 & 9,106  & 1.4505 & 0.16 \\
-24 & -       & -      & -      & -      & -    & -     & -    & -    & 192    & 1.5329 & 0.12 \\
\tableline
All* & 130,602 & 1.4432 & 1.4437 & 1.4443 & 0.50 & 0.090 & 1.05 & 0.25 & - & - & - \\
\enddata
\label{tab:lumbins}
\tablecomments{Properties of the ${\rm M}_{r,0.1}$ luminosity bins.  Columns 2 through 9 give values for volume limited magnitude bins while columns 10, 11 and 12 give values for all sample galaxies within the luminosity range.  $\overline{\rm ap}$ is the mean aperture fraction.  * The sample as a whole is not volume limited but the more detailed information is given for reference.}

\end{deluxetable}

\begin{deluxetable}{cccccl}
\tabletypesize{\scriptsize}
\tablecaption{Multi-burst SFH Models}
\tablewidth{0pt}
\tablehead{
\colhead{Name} & \colhead{Type} & \colhead{Length} &
\colhead{Relative} & \colhead{Spacing} & \colhead{Underlying}\\
\colhead{} & \colhead{} & \colhead{(Myr)} &
\colhead{Strength} & \colhead{(Gyr)} & \colhead{SFH}}

\startdata
SFH 1 & burst  & 200 & -   & 3   & none  \\
SFH 2 & burst  & 200 & -   & 1   & none  \\
SFH 3 & burst  & 200 & 4.0 & 3   & constant  \\
SFH 4 & gasp   & 200 & 0.0 & 1.5 & constant  \\
SFH 5 & gasp   & 200 & 0.1 & 1.5 & exponential  \\
SFH 6 & burst   & 200 & 5.0 & 1.5 & exponential  \\
\enddata
\label{tab:sfhs}
\end{deluxetable}

\clearpage

\clearpage

\end{document}